\documentclass[aps,prd,twocolumn,nofootinbib,preprintnumbers,10pt]{revtex4-2}
\usepackage{style}

\begin{document}

\title{Electronic Direct Detection of Light Dark Matter with Intermediate-Mass Mediators}

\author{Connor Stratman\,\orcidlink{0009-0006-3574-866X}}
\email{connor35@illinois.edu}
\affiliation{Department of Physics, Grainger College of Engineering, University of Illinois Urbana-Champaign, Urbana, IL 61801, USA}

\author{Tanner Trickle\,\orcidlink{0000-0003-1371-4988}}
\email{ttrickle@illinois.edu}
\affiliation{Department of Physics, Grainger College of Engineering, University of Illinois Urbana-Champaign, Urbana, IL 61801, USA}

\begin{abstract}

Recent years have seen dramatic improvements in the sensitivity of electron-based direct detection experiments. Typically, the sensitivity to dark matter scattering is determined in the light and heavy mediator mass limits. In this paper we show that the light and heavy mediator mass limits are not separated by a single scale, but instead can be separated by up to three orders of magnitude in mediator mass for sub-GeV mass dark matter. We calculate the background-free sensitivity in Si and Ge targets, and a projected DAMIC-M sensitivity, to sub-GeV mass dark matter models with ``intermediate-mass" mediators between the light and heavy mediator limits. This allows us to determine the precise range of mediator masses that electron-based direct detection experiments are sensitive to when the dark matter relic abundance is generated via freeze-in. We make the calculations presented here publicly available in an updated release of~\EXCEEDDM. 

\end{abstract}

\date{\today}
\maketitle

\tableofcontents

\section{Introduction}
\label{sec:introduction}

Direct detection experiments are rapidly improving their reach to light, sub-GeV mass dark matter (DM). Leading these searches are electron-based direct detection experiments using semiconducting targets, whose eV-scale band gaps allow them to achieve sensitivity down to MeV-scale DM masses. Current experiments using either silicon (Si) or germanium (Ge) targets include CDEX~\cite{CDEX:2019exx,CDEX:2022kcd}, DAMIC~\cite{DAMIC-M:2023gxo,DAMIC-M:2025luv}, EDELWEISS~\cite{EDELWEISS:2020fxc}, SENSEI~\cite{SENSEI:2023zdf}, and SuperCDMS~\cite{SuperCDMS:2019jxx,SuperCDMS:2020ymb}. Building upon these efforts, the Oscura~\cite{Oscura:2022vmi} experiment has been proposed as a next-generation, large-scale successor. Recently, the DAMIC experiment at the Modane Underground Laboratory (DAMIC-M) reached an important milestone. In Ref.~\cite{DAMIC-M:2025luv} DAMIC-M was the first experiment to exclude freeze-in produced DM that interacts with the Standard Model via a light kinetically-mixed dark photon~\cite{Hall:2009bx,Chu:2011be,Hambye:2019dwd,Dvorkin:2019zdi} for some DM masses in the MeV - GeV mass range. This is significant because freeze-in produced DM is one of the main theoretically well-motivated benchmark models these experiments were built to probe.

The DAMIC-M results motivate further questions about what DM models these experiments are sensitive to. Here we study the canonical light DM model, where DM freezes-in from the Standard Model via a kinetically-mixed dark photon (defined in detail in Sec.~\ref{subsec:interaction_lagrangian}), but our focus is on determining how the detectability depends on the dark photon mass. Usually, the DM-electron scattering rate in direct detection experiments is computed in two limits of the dark photon mass, the ``light" and ``heavy" mediator limits; see Refs.~\cite{Lin:2019uvt,Kahn:2021ttr,Zurek:2024qfm} for recent reviews. While DAMIC-M is sensitive to freeze-in produced DM in the light mediator limit, detectability drops sharply as the dark photon mass increases. This means that, for freeze-in produced DM, there is a maximum dark photon mass DAMIC-M, or any other direct detection experiment, is sensitive to. Our goal here is to precisely understand the detectability of this target benchmark DM model for dark photon masses between the light and heavy mediator limits, which we refer to as the ``intermediate-mass mediator" regime. To do this, we perform novel DM-electron scattering rate calculations in Si and Ge targets that explicitly account for the mediator mass. All calculations presented here are made publicly available in an updated release of \EXCEEDDM, and the results are available at Ref.~\cite{stratman_2026_19805613}.

This paper is organized as follows. In Sec.~\ref{sec:formalism} we review the canonical light DM interaction Lagrangian, coupled to the Standard Model via a kinetically-mixed dark photon, and the resulting DM-electron scattering rate calculation. In Sec.~\ref{sec:intermediate_mass_mediators} we discuss and define the intermediate-mass mediator regime in the context of electron-based direct detection. In Sec.~\ref{sec:sensitivity} we calculate the sensitivity of current and future electron-based direct detection experiments to the freeze-in benchmark target at dark photon masses in the intermediate-mass regime. Specifically, in Sec.~\ref{subsec:background-free_sensitivity} we calculate the sensitivity of background-free experiments using Si and Ge targets, and in Sec.~\ref{sec:projected_DAMIC_sensitivity} we project the sensitivity of DAMIC-M~\cite{DAMIC-M:2025luv} using a simplified statistical procedure that includes backgrounds. 

\section{Formalism}
\label{sec:formalism}

We begin with a discussion of the formalism used throughout this paper. In Sec.~\ref{subsec:interaction_lagrangian} we define the canonical light DM model, interacting with the Standard Model via a kinetically-mixed dark photon, and in Sec.~\ref{subsec:dm_electron_scattering_rate} we review the DM-electron scattering rate calculation with an emphasis on where the dark photon mass appears.

\subsection{Interaction Lagrangian}
\label{subsec:interaction_lagrangian}

The light DM model considered here features a Dirac fermion DM candidate, $\chi$, charged under a broken $U(1)'$ gauge symmetry. The DM interacts with the Standard Model through a kinetically-mixed massive dark photon, $A'$; the relevant Lagrangian below the electroweak scale given by,
\begin{align}
    \mathcal{L} \supset \bar{\chi} \left( i \slashed{D}' - m_\chi \right) \chi  + \frac{1}{2} m_{A'}^2 A^{\prime\,\mu} A'_\mu - \frac{\varepsilon}{2} F^{\mu \nu} F'_{\mu \nu} \, ,
    \label{eq:L}
\end{align}
where $m_\chi$ and $m_{A'}$ are the DM and dark photon mass, respectively, $D'_\mu = \partial_\mu + i g' A'_\mu$ is the dark gauge covariant derivative with gauge coupling $g'$, $F_{\mu\nu}^{(\prime)}$ is the (dark) electromagnetic field strength tensor and $\varepsilon$ is the kinetic mixing parameter. We have ignored the dark photon kinetic term for simplicity. When the dark photon is massive it is convenient to work in the photon-dark photon mass basis, which is found with the field redefinition $A_\mu \rightarrow A_\mu - \varepsilon A'_\mu$ when $\varepsilon \ll 1$. After this field redefinition, all electromagnetically charged Standard Model particles acquire a coupling to the dark photon, and the coupling relevant to electron-based direct detection is,
\begin{align}
    \mathcal{L} \supset A'_\mu \left( - g' \, \bar{\chi} \gamma^\mu \chi + \varepsilon e \; \bar{\ell} \gamma^\mu \ell \right) \, ,
    \label{eq:L_int}
\end{align}
where $e = -|e|$ is the electromagnetic charge, following the convention in Ref.~\cite{Peskin:1995ev}, and $\ell$ is the electron field. 

From Eq.~\eqref{eq:L_int} we see that $\varepsilon g'$ is the combination of couplings that will determine the detectability of $\chi$ in direct detection experiments. However, it is conventional to use a reference cross section, $\bar{\sigma}_e$, instead, defined as~\cite{Griffin:2019mvc,Kahn:2021ttr},
\begin{align}
    \bar{\sigma}_e \equiv \frac{\mu_{\chi e}^2}{\pi} \frac{g'^2 \, (\varepsilon e)^2}{\left( \alpha^2 m_e^2 + m_{A'}^2 \right)^2} \, ,
    \label{eq:cross_section_definition}
\end{align}
where $m_e$ is the electron mass, $\mu_{\chi e} = m_\chi m_e / (m_\chi + m_e)$ is the reduced DM-electron mass, and $\alpha = e^2 / 4 \pi$ is the fine-structure constant. Alternative definitions of $\bar{\sigma}_e$ commonly used throughout the literature are compared in App.~\ref{app:reference_cross_section_definitions}.

\subsection{Dark Matter-Electron Scattering Rate}
\label{subsec:dm_electron_scattering_rate}

Given the interaction Lagrangian in Eq.~\eqref{eq:L_int}, the DM-electron scattering rate is given by~\cite{Griffin:2021znd,Trickle:2022fwt,Dreyer:2023ovn}
\begin{align}
    R = \frac{\pi \bar{\sigma}_e \rho_\chi}{V \mu_{\chi e}^2 m_\chi \rho_T} \sum_{IF} \int \frac{\dd^3 \mathbf{q}}{(2 \pi)^3} \mathcal{F}^2(q) g(\mathbf{q},\omega) \frac{|\langle F | e^{i \mathbf{q} \cdot \mathbf{x}} | I \rangle|^2}{|\varepsilon(q,\omega)|^2} ,
    \label{eq:scattering_rate}
\end{align}
where $V$ is the target volume, $\rho_T$ is the target mass density, $\rho_\chi \approx 0.4 \, \text{GeV} / \text{cm}^3$ is the local DM density, $|I\rangle,|F\rangle$ are the initial (filled) and final (empty) target electron states, $\mathbf{q}$ is the momentum deposited to the target, $q \equiv |\mathbf{q}|$, and $\omega = E_F - E_I$ is the energy deposited. Screening effects, due to in-medium dark photon-photon mixing~\cite{Knapen:2021run,Hochberg:2021pkt,Griffin:2021znd,Trickle:2022fwt}, are incorporated with the dielectric function, $\varepsilon(q, \omega)$, which we assume to be isotropic since our focus is on Si and Ge targets. We note that Eq.~\eqref{eq:scattering_rate} can also be rewritten in terms of the energy loss function~\cite{Hochberg:2021pkt,Knapen:2021run}, and we do not include local field effects~\cite{Knapen:2021run,Dreyer:2026bmz}.

The kinematic function $g(\mathbf{q},\omega)$ encodes the dynamics of the DM halo~\cite{Trickle:2019nya}
\begin{align}
    g(\mathbf{q},\omega) = 2 \pi \int \dd^3 \mathbf{v}  \, f_\chi(\mathbf{v}) \, \delta ( \omega - \omega_\mathbf{q}) \,,
\end{align}
where $f_\chi(\mathbf{v})$ is the lab-frame DM velocity distribution, and 
\begin{align}
    \omega_\mathbf{q} = \mathbf{q} \cdot \mathbf{v} - \frac{q^2}{2 m_\chi} \, ,
    \label{eq:omega_q}
\end{align}
is the energy deposited in a scattering event. We adopt a boosted, truncated Maxwell-Boltzmann distribution~\cite{Kahn:2021ttr,Baxter:2021pqo} for $f_\chi(\mathbf{v})$,
\begin{align}
    f_{\chi}(\mathbf{v}) \propto \exp\left( {\frac{-|\mathbf{v} + \mathbf{v}_\text{e}|^2}{v_0^2}} \right)\Theta(v_{\textrm{esc}} - |\mathbf{v} + \mathbf{v}_\text{e}|) \,,
    \label{eq:DM_velocity_distribution}
\end{align}
whose overall normalization is set by requiring $\int \dd^3\mathbf{v} \, f_\chi(\mathbf{v}) = 1$. Following the recommended conventions in Ref.~\cite{Baxter:2021pqo}, we assume $v_0 = 238$ km/s, $v_{\textrm{esc}}=544$ km/s and $|\mathbf{v}_\text{e}| = v_e =250$ km/s.

Finally, the central quantity for this work is the mediator form factor
\begin{align}
    \mathcal{F}(q) \equiv \frac{\alpha^2 m_e^2 + m_{A'}^2}{q^2 + m_{A'}^2} \,,
    \label{eq:exact_form_factor}
\end{align}
which encapsulates the mediator mass dependence of the DM-electron scattering rate, a relationship we explore in detail next.

\begin{figure}[ht]
    \centering
    \includegraphics[width=\linewidth]{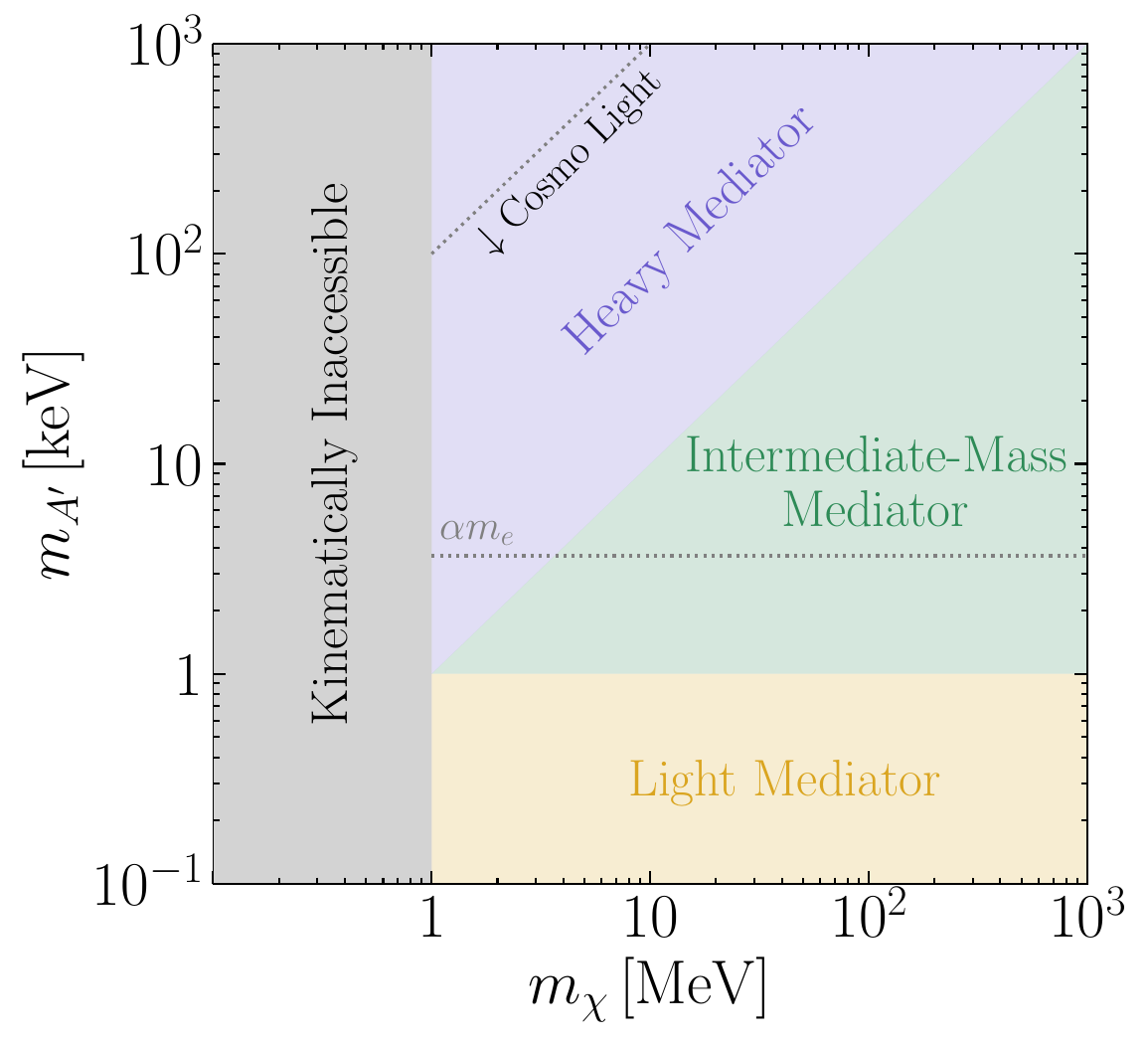}
    \caption{Illustration of the different dark photon mass limits relevant for eV-scale threshold, $\omega_\text{th} \sim \text{eV}$, direct detection experiments. When the dark photon is much lighter than the minimum momentum transfer, $\sim \omega_\text{th} / v$, where $v \sim 10^{-3}$ is the DM velocity, the usual light mediator limit applies (yellow region, labeled ``Light Mediator"). When the dark photon is much heavier than the maximum momentum transfer, $\sim m_\chi v$, the usual heavy mediator limit applies (purple region, labeled ``Heavy Mediator"). In between these two regimes is the \textit{intermediate-mass mediator} regime where the exact form factor in Eq.~\eqref{eq:exact_form_factor} must be used, since momentum larger or smaller than the dark photon mass can be transferred to the target (green region, labeled ``Intermediate-Mass Mediator"). DM with $m_\chi \lesssim \text{MeV}$ (gray region, labeled ``Kinematically Inaccessible") does not have enough kinetic energy to excite an electron over an eV-scale threshold. Lastly, for $m_{A'}\lesssim m_\chi / 10$ (below the dotted black line, labeled ``Cosmo Light") the dark photon is effectively massless for all cosmological processes relevant for DM freeze-in~\cite{Chu:2011be,Hambye:2019dwd}.}
    \label{fig:mA_vs_m_chi_regions}
\end{figure}

\section{Intermediate-Mass Mediator}
\label{sec:intermediate_mass_mediators}

The DM-electron scattering rate, Eq.~\eqref{eq:scattering_rate}, is typically computed in two limits of the mediating dark photon mass: the ``light" and ``heavy" mediator limits~\cite{Essig:2011nj,Essig:2015cda,Griffin:2021znd}. Practically these correspond to the limits where the mediator form factor, Eq.~\eqref{eq:exact_form_factor}, reduces to a simple form. The light mediator limit corresponds to $\mathcal{F} \approx \alpha^2 m_e^2 / q^2$, while the heavy mediator limit corresponds to $\mathcal{F} \approx 1$. It is often assumed that the cutoff scale in $m_{A'}$ between these regimes is simply $\alpha \, m_e$, since this would be the typical momentum transferred if the DM was heavier than an electron, $m_\chi \gg m_e$, and scattered off of a free electron with velocity $\alpha$~\cite{Essig:2011nj,Lin:2019uvt}. However electrons inside semiconducting targets are not free particles, which changes DM-electron scattering kinematics significantly. What physically matters is how $m_{A'}$ compares to the momentum transferred to the target.\footnote{Additionally, note that while both the reference DM-electron cross section in Eq.~\eqref{eq:cross_section_definition}, and the mediator form factor in Eq.~\eqref{eq:exact_form_factor}, depend on the combination of $\alpha^2 m_e^2 + m_{A'}^2$, the total DM-electron scattering rate, Eq.~\eqref{eq:scattering_rate}, does \textit{not} depend on the factors of $\alpha^2 m_e^2 + m_{A'}^2$. That is, the total DM-electron scattering rate with fixed couplings is \textit{independent} of the combination $\alpha^2 m_e^2 + m_{A'}^2$; its presence is purely conventional.} The light mediator limit ($\mathcal{F} \propto 1 / q^2$) then corresponds to the limit where $m_{A'}$ is much smaller than the minimum momentum transferred to the target, and the heavy mediator limit ($\mathcal{F} \approx 1$) corresponds to the limit where $m_{A'}$ is much larger than the maximum momentum transferred to the target. Since for a general target the minimum momentum transfer is not equal to the maximum, there is an intermediate regime. When $m_{A'}$ is comparable to the momentum transferred in this intermediate momentum regime, we refer to this as the ``intermediate-mass mediator" regime. 

To more rigorously define the intermediate-mass mediator regime we must discuss DM-electron scattering kinematics and the minimum, and maximum, kinematically allowed momentum transfers. The energy deposited in a scattering event, Eq.~\eqref{eq:omega_q}, is bounded by~\cite{Trickle:2019nya},
\begin{align}
    \omega_\mathbf{q} \leq q \,v_\text{max} - \frac{q^2}{2 m_\chi} \, ,
\end{align}
where $v_\text{max} = v_\text{esc} + v_\text{e}$ is the maximum DM velocity in the detector frame. Requiring that $\omega \geq \omega_\text{th}$, where $\omega_\text{th}$ is the threshold energy of the detector, the momentum transfer is then bounded between $q_- < q < q_+$, where
\begin{align}
    \label{eq:exact_momentum_bounds}
    q_\pm = m_\chi v_\text{max} \pm \sqrt{ (m_\chi v_\text{max})^2 - 2 m_\chi \omega_\text{th}} \, .
\end{align}
Furthermore, $q_\pm$ can themselves be bounded by simpler forms,
\begin{align}
    \frac{\omega_\text{th}}{v_\text{max}} < q_- < q < q_+ < 2 m_\chi v_\text{max} \, .
    \label{eq:momentum_scales}
\end{align}
To understand the scale of these momentum transfers, note that current electron-based direct detection experiments operate with $\omega_\text{th} \sim \text{eV}$. Therefore the minimum and maximum momentum scales are $\omega_\text{th} / v_\text{max} \sim \text{keV}$ and $m_\chi v_\text{max} \sim \text{keV} \, (m_\chi / \text{MeV})$, respectively. This range can be as large as three orders of magnitude for GeV mass DM. Comparing $m_{A'}$ to the momentum scales in Eq.~\eqref{eq:momentum_scales} allows us to summarize the light, heavy, and intermediate-mass regimes as, 
\begin{align}
    m_{A'} &\ll \frac{\omega_\text{th}}{v_\text{max}} & \text{Light Mediator} \phantom{\, .}\nonumber\\ 
    \frac{\omega_\text{th}}{v_\text{max}} \lesssim m_{A'} &\lesssim 2 m_\chi v_\text{max} & \text{Intermediate-Mass Mediator}\phantom{\, .}  \nonumber \\ 
    2 m_\chi v_\text{max} &\ll m_{A'} & \text{Heavy Mediator} \, .
    \label{eq:regimes}
\end{align}

Fig.~\ref{fig:mA_vs_m_chi_regions} illustrates the light, heavy, and intermediate-mass mediator regimes in the $m_{A'}-m_\chi$ plane. For simplicity, we assume a DM velocity of $v \sim 10^{-3}$ and $\omega_\text{th} \sim 1 \, \text{eV}$. The cutoff between the heavy mediator (purple) and intermediate-mass mediator (green) regimes is set by $m_\chi v$, and the cutoff between the light mediator (yellow) and intermediate-mass mediator regimes is set by $\omega_\text{th} / v$. The gray region is kinematically inaccessible, since the DM cannot deposit enough energy to excite an electron across the band gap, and we also show the $\alpha m_e$ scale (dotted gray) for comparison. 

For $m_{A'} \lesssim m_\chi / 10$, the dark photon is always ``cosmologically" light. That is, the dark photon can be considered massless from the perspective of DM freeze-in production in the early universe. Strictly speaking this is the case when $m_{A'}$ is much smaller than the Standard Model plasma frequency, $\omega_\text{pl} \sim e \, T$, where $T$ is the Standard Model temperature, such that on-shell dark photon production is screened~\cite{Chu:2011be,Hambye:2019dwd}. Since most freeze-in DM production occurs when $T \sim m_\chi$, the relevant cutoff is $m_{A'} \ll e \, m_\chi$, which we approximate as $m_\chi / 10$ in Fig.~\ref{fig:mA_vs_m_chi_regions} for illustration (with a boundary labeled by ``Cosmo Light").

The distinction between dark photon mass scales in cosmological versus direct detection contexts is important, as the former fixes the Lagrangian couplings ($\varepsilon g'$) required to generate the relic abundance. Since our focus here is on the intermediate-mass regime relevant for direct detection, all the dark photon masses we consider are well in the cosmologically light regime. In this cosmologically light regime the $\varepsilon g'$ required to produce the DM relic abundance, calculated in detail in Ref.~\cite{Bhattiprolu:2023akk}, is \textit{independent} of $m_{A'}$. All $m_{A'}$ dependence in the freeze-in target reference cross section, $\bar{\sigma}_e^\text{FI}$, calculated by substituting the required $\varepsilon g'$ into Eq.~\eqref{eq:cross_section_definition}, is due to the definition of Eq.~\eqref{eq:cross_section_definition}. Specifically, this means that $\bar{\sigma}_e^\text{FI} \propto m_{A'}^0$ when $m_{A'} \ll \alpha m_e$, and $\bar{\sigma}_e^\text{FI} \propto 1 / m_{A'}^4$ when $m_{A'} \gg \alpha m_e$. This will be important in Sec.~\ref{sec:sensitivity}, where we assess whether current and future electron-based direct detection experiments can reach the sensitivity required to probe $\bar{\sigma}_e^\text{FI}$-sized cross sections as a function of $m_{A'}$.

\begin{figure*}[ht!]
    \centering
    \includegraphics[width=0.45\linewidth]{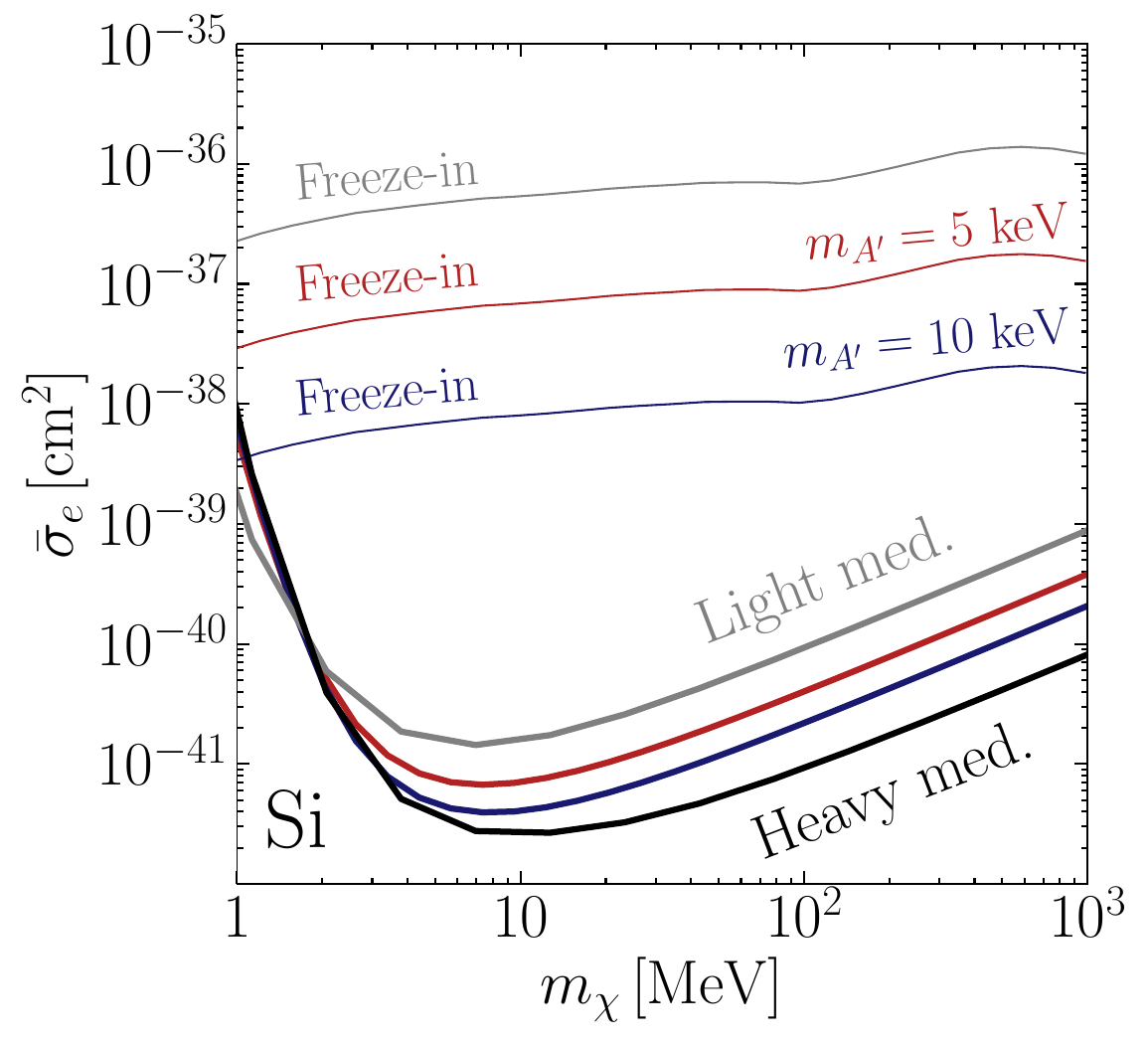}
    \includegraphics[width=0.45\linewidth]{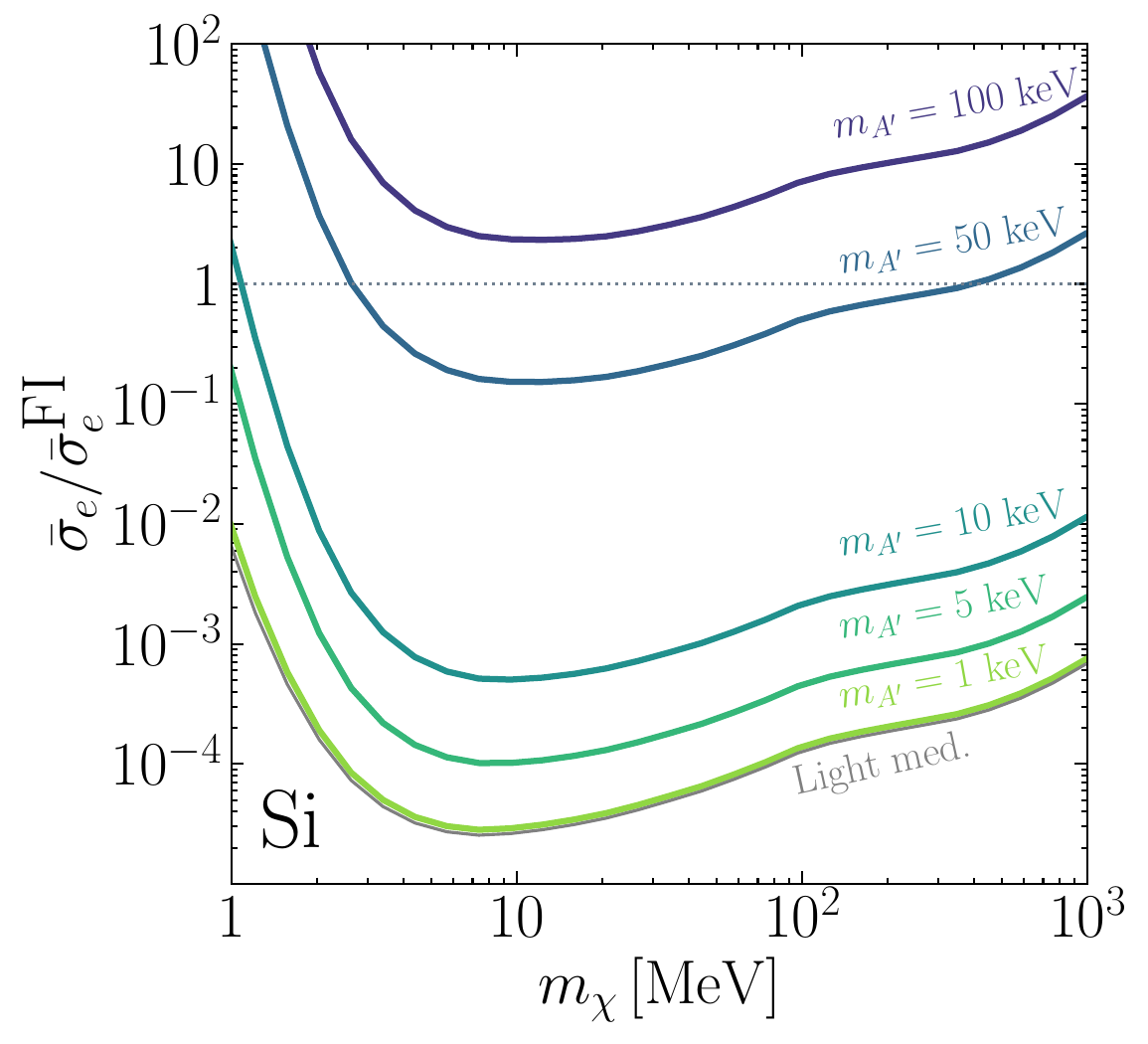}
    \includegraphics[width=0.45\linewidth]{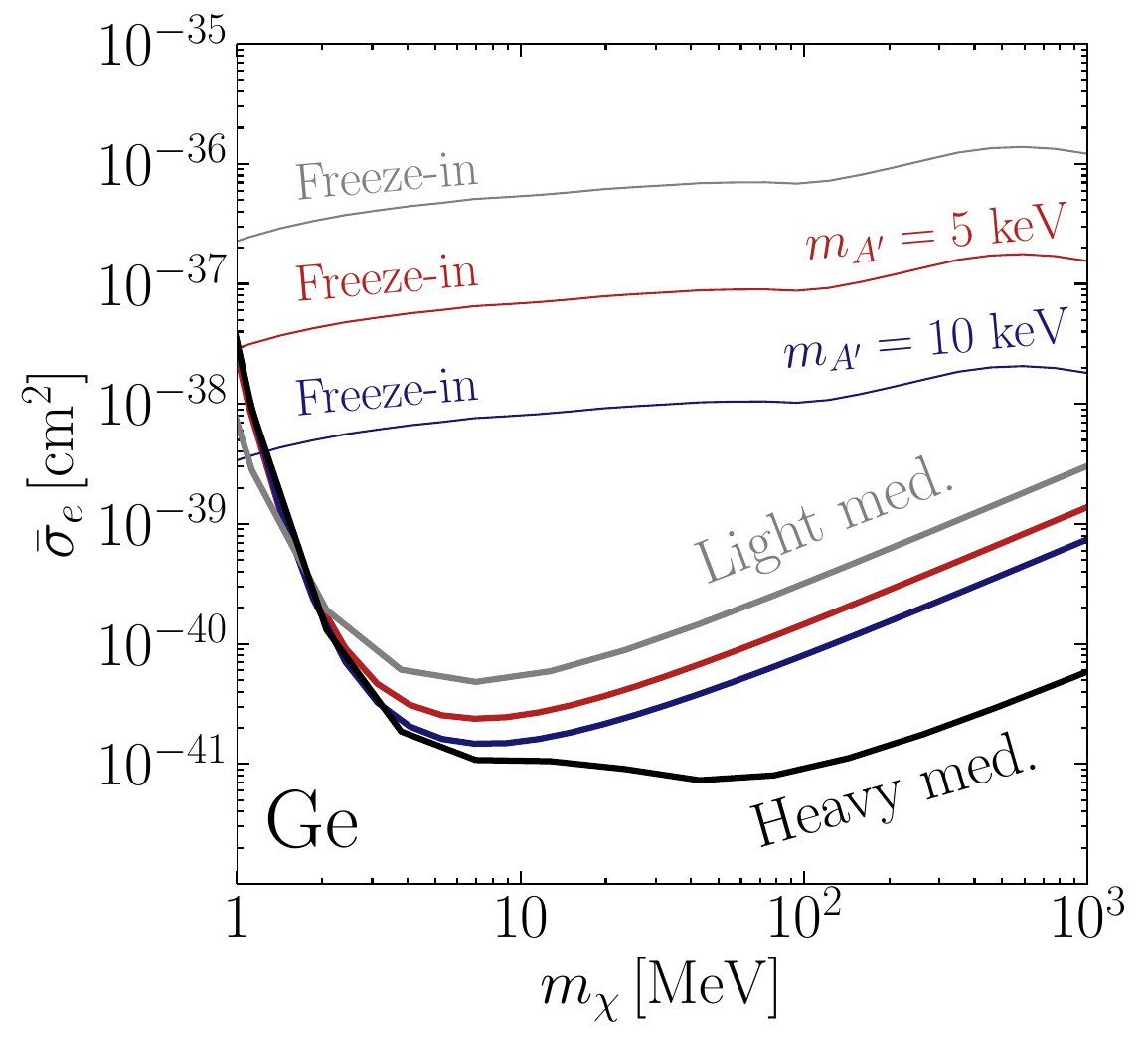}
    \includegraphics[width=0.45\linewidth]{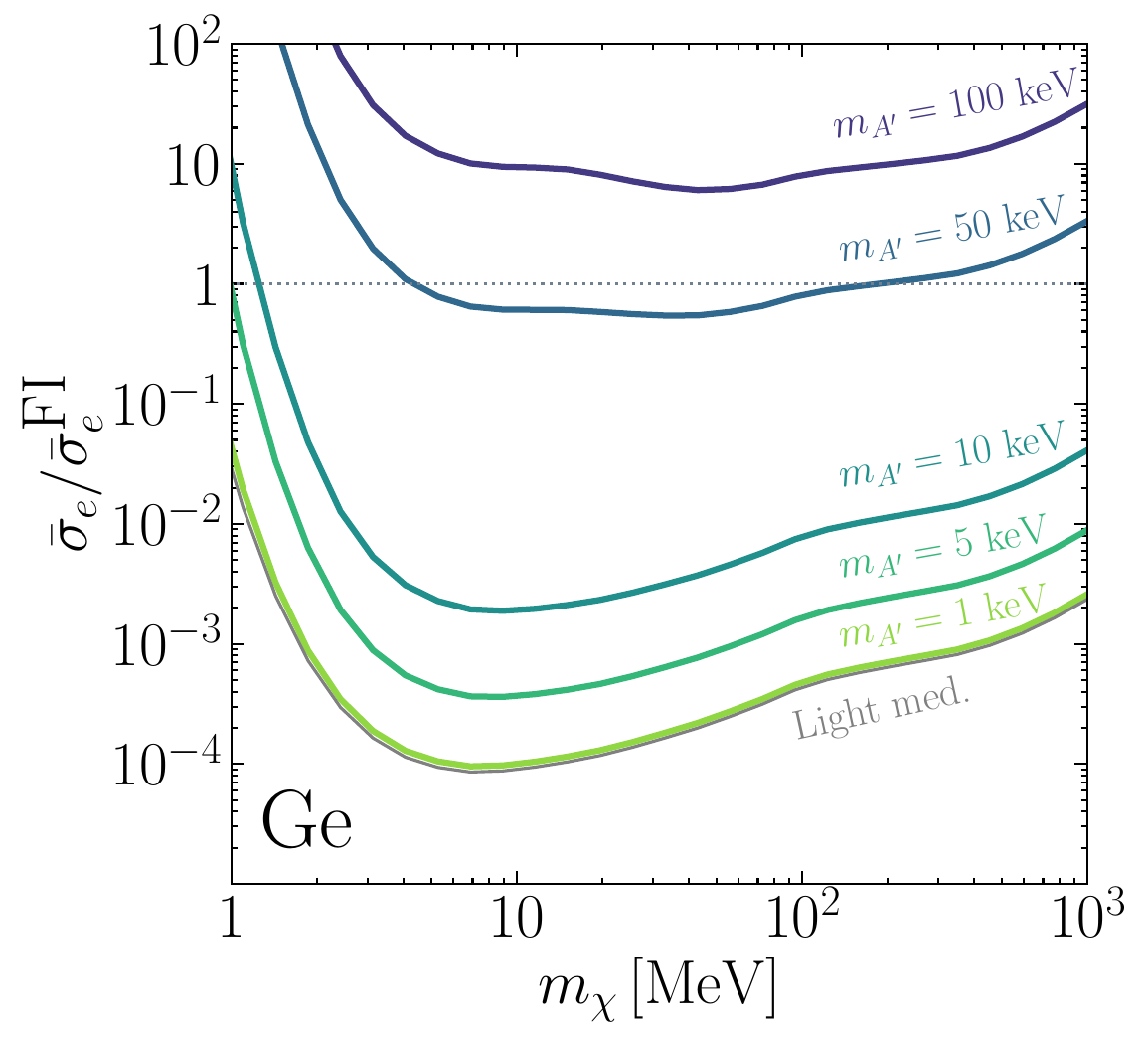}
    \caption{Projected 95\% C.L. cross section sensitivity for background-free Si (\textbf{top}) and Ge (\textbf{bottom}) detectors, assuming an ionization charge threshold $Q \geq 1$ and kg-yr exposure. \textbf{Left:} Comparison of the cross section sensitivity (thick lines) to the freeze-in cross section, $\bar{\sigma}_e^{\textrm{FI}}$ (thin lines, Ref.~\cite{Bhattiprolu:2023akk}), for different dark photon masses, $m_{A'}$. Gray lines apply to dark photons in the light mediator limit (Sec.~\ref{sec:intermediate_mass_mediators}), red lines for $m_{A'} = 5 \, \text{keV}$, and blue lines for $m_{A'} = 10 \, \text{keV}$. The cross section sensitivity in the light (gray) and heavy (black) mediator limits are taken from Ref.~\cite{Trickle:2022fwt}. \textbf{Right:} The ratio of the cross section sensitivity to the freeze-in cross section for a selection of dark photon masses. These colored lines indicate the detectability of a DM model: if $\bar{\sigma}_e / \bar{\sigma}_e^\text{FI} < 1$ then an experiment is sensitive to those DM and dark photon masses with freeze-in sized cross sections. A dashed gray line at $\bar{\sigma}_e / \bar{\sigma}_e^\text{FI} = 1$ is shown for reference.}
    \label{fig:bkg_free_cs_sensitivity_mX}
\end{figure*}

\section{Electron-Based Direct Detection Sensitivity}
\label{sec:sensitivity}

We now present the projected sensitivity of electron-based direct detection experiments to the DM model in Sec.~\ref{subsec:interaction_lagrangian}, explicitly accounting for the dark photon mass. In Sec.~\ref{subsec:background-free_sensitivity}, we focus on background-free Si and Ge targets, and compute the cross section sensitivity as a function of both $m_\chi$ and $m_{A'}$, for $m_{A'}$ in the intermediate-mass regime. Additionally we show the sensitivity to $m_{A'}$ as a function of $m_\chi$ for freeze-in sized cross sections. In Sec.~\ref{sec:projected_DAMIC_sensitivity}, we use a simplified statistical procedure that accounts for backgrounds to compute the projected cross section sensitivity of DAMIC-M~\cite{DAMIC-M:2025luv} as a function of both $m_\chi$ and $m_{A'}$. We then show the projected sensitivity of DAMIC-M to $m_\chi$ and $m_{A'}$ for freeze-in sized cross sections.

All calculations use electronic configuration files and numerically computed dielectric functions that are available in Ref.~\cite{trickle_2022_zenodo}. Additionally, the calculations done here have been added to a new version of \EXCEEDDM \, and have been made publicly available at Ref.~\cite{stratman_2026_19805613}.

\subsection{Background-Free Sensitivity}
\label{subsec:background-free_sensitivity}

Starting from the DM-electron scattering rate in Eq.~\eqref{eq:scattering_rate}, we calculate the 95\% confidence level (C.L.) cross section assuming a kg-yr exposure and no backgrounds. We calculate the cross section sensitivity for specific dark photon masses by including the explicit mediator mass dependence in the DM-electron scattering rate via the form factor, $\mathcal{F}(q)$, Eq.~\eqref{eq:exact_form_factor}.

In the left panels of Fig.~\ref{fig:bkg_free_cs_sensitivity_mX} we plot cross section sensitivities for Si (top) and Ge (bottom) detectors in the standard light mediator and heavy mediator limits, as well for two mediator masses in the intermediate-mass regime. To understand the sensitivity of this background-free experiment to freeze-in produced DM, we also plot the freeze-in target reference cross section, $\bar{\sigma}_e^{\textrm{FI}}$, (thin lines) calculated by the procedure described in Sec.~\ref{sec:intermediate_mass_mediators}. For a given $m_\chi$ and $m_{A'}$, if $\bar{\sigma}_e < \bar{\sigma}_e^{\textrm{FI}}$, the detector is sensitive to freeze-in produced DM.

An alternative comparison between the cross section sensitivity and $\bar{\sigma}_e^\text{FI}$ is to plot their ratio directly. In the right panels of Fig.~\ref{fig:bkg_free_cs_sensitivity_mX}, we plot this ratio as a function of $m_\chi$ for a selection of dark photon masses with Si (top) and Ge (bottom) detectors. Similar to the left panels, the detector is sensitive to freeze-in produced DM when $\bar{\sigma}_e / \bar{\sigma}_e^{\textrm{FI}} < 1$. We do not plot the heavy mediator limit in the right panels because $\bar{\sigma}_e^\text{FI}$ becomes prohibitively small at large mediator masses due to the scaling $\bar{\sigma}_e^{\textrm{FI}} \propto 1/m_{A'}^4$.

From Fig.~\ref{fig:bkg_free_cs_sensitivity_mX} it is clear that there is a non-trivial dependence on both $m_{A'}$ and $m_\chi$ in determining the sensitivity of a detector to freeze-in produced DM. For a Si target (Fig.~\ref{fig:bkg_free_cs_sensitivity_mX}, top right panel) the shape of the cross section ratio is similar across all mediator masses shown. For a Ge target (Fig.~\ref{fig:bkg_free_cs_sensitivity_mX}, bottom right panel) larger dark photon masses, $m_{A'} \gtrsim 10$ keV, shift the minimum of $\bar{\sigma}_e / \bar{\sigma}_e^{\textrm{FI}}$ to larger $m_\chi$. This can be understood from the enhancement of the scattering rate due to high momentum transfers in Ge, relevant for $m_\chi \gtrsim 10$ MeV~\cite{Griffin:2021znd,Dreyer:2023ovn,Dreyer:2026bmz}. Since $m_{A'}$ sets the scale of the important momentum transfers, increasing the mediator mass allows large $q$ transitions in Ge to dominate the rate. 

\begin{figure*}[ht]
    \centering
    \includegraphics[width=0.45\linewidth]{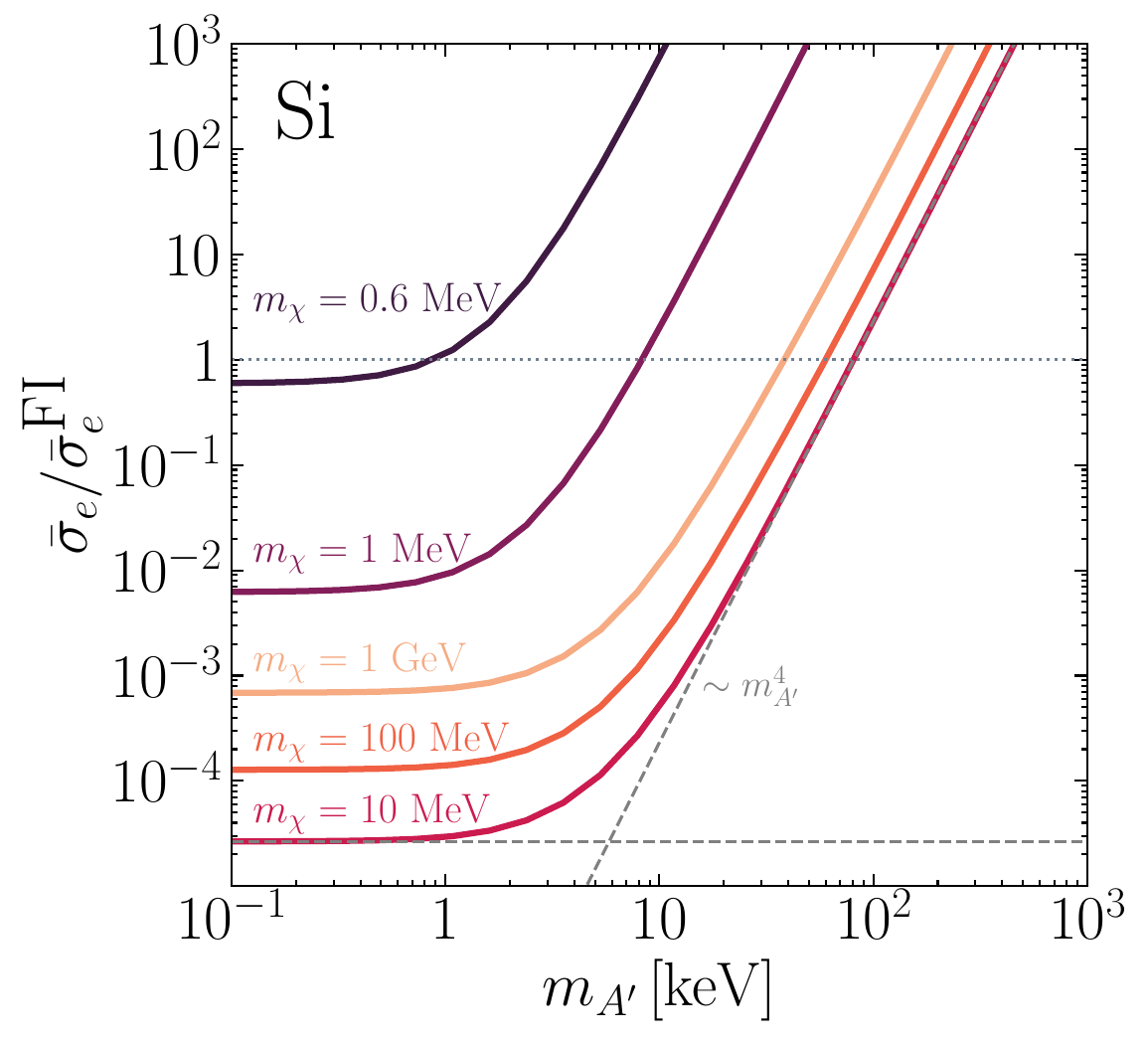}
    \includegraphics[width=0.45\linewidth]{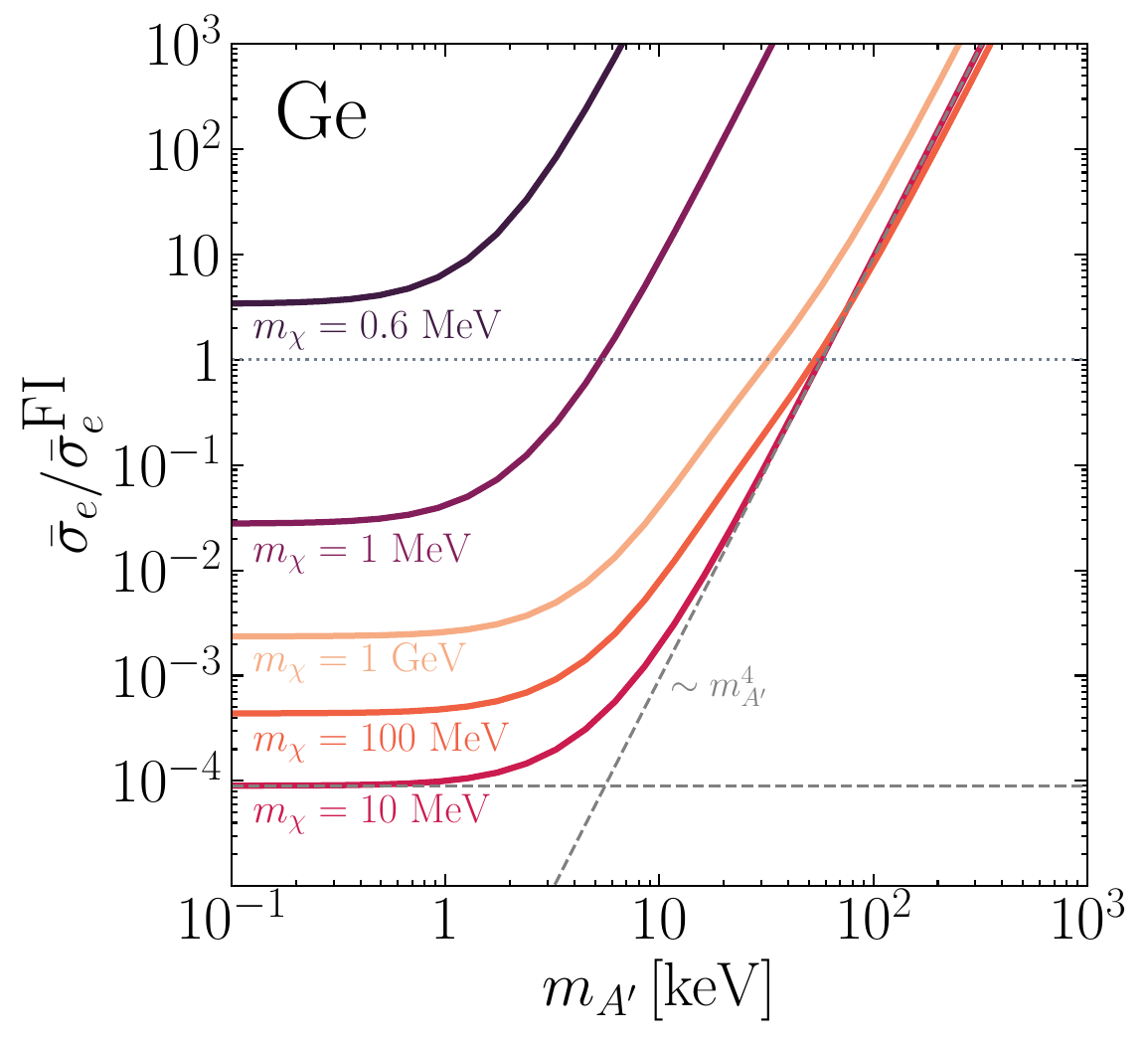}
    \caption{The ratio of the projected 95\% C.L. cross section sensitivity to the freeze-in cross section, $\bar{\sigma}_e / \bar{\sigma}_e^{\textrm{FI}}$, for background-free Si (\textbf{left}) and Ge (\textbf{right}) detectors, assuming an ionization charge threshold $Q \geq 1$ and kg-yr exposure, as a function of dark photon mass, $m_{A'}$. The different colored lines correspond to selected DM masses, $m_\chi$, and, similar to Fig.~\ref{fig:bkg_free_cs_sensitivity_mX}, the dotted gray line indicates the boundary of detectability. Dashed gray lines illustrate the expected scaling in the light ($\bar{\sigma}_e / \bar{\sigma}_e^\text{FI} \propto m_{A'}^0$) and heavy ($\bar{\sigma}_e / \bar{\sigma}_e^\text{FI} \propto m_{A'}^4$) mediator limits. For $m_\chi \gtrsim 10$ MeV, in Ge, the effect of intermediate dark photon masses on $\bar{\sigma}_e / \bar{\sigma}_e^\text{FI}$ is more pronounced than in Si, as seen by the approximate scaling of $\bar{\sigma}_e / \bar{\sigma}_e^{\textrm{FI}} \propto m_{A'}^3$ near $m_{A'} \sim 20 \, \text{keV}$. This is due to the enhancement of the scattering rate from higher momentum transfers in Ge~\cite{Griffin:2021znd,Dreyer:2023ovn,Dreyer:2026bmz}.}
    \label{fig:bkg_free_cs_sensitivity_mA}
\end{figure*}

To further explore the dependence of $\bar{\sigma}_e / \bar{\sigma}_e^\text{FI}$ on the dark photon mass, in Fig.~\ref{fig:bkg_free_cs_sensitivity_mA} we fix the DM mass and plot the cross section ratio as a function of dark photon mass. For both Si and Ge targets there are clear scalings of the sensitivity to freeze-in target ratio, $\bar{\sigma}_e / \bar{\sigma}_e^\text{FI}$, as a function of $m_{A'}$ in the light and heavy mediator limits, which we will discuss in detail now.

To understand the $m_{A'}$ scaling in the light and heavy mediator limits note that, due to the definition of the mediator form factor $\mathcal{F}(q)$ in Eq.~\eqref{eq:exact_form_factor}, the cross section sensitivity is independent of $m_{A'}$ in the light and heavy mediator limits. Therefore the $\bar{\sigma}_e / \bar{\sigma}^\text{FI}_e$ scaling in the light and heavy mediator limits is determined by $\bar{\sigma}^\text{FI}_e$ alone. In the light mediator regime, $\bar{\sigma}_e^\text{FI} \sim m_{A'}^0$, while in the heavy mediator regime $\bar{\sigma}_e^{\textrm{FI}} \propto 1 / m_{A'}^{4}$. We show both these scaling limits as dashed gray lines for $m_\chi=10$ MeV in Fig.~\ref{fig:bkg_free_cs_sensitivity_mA}. These scaling limits allow us to discern the approximate intermediate-mass regime for each $m_\chi$ shown. For example, in a Si target, the intermediate mass regime for $m_\chi=10$ MeV is approximately $1 \, \text{keV} \lesssim m_{A'} \lesssim 30 \, \text{keV}$, in close agreement with our estimate in Fig.~\ref{fig:mA_vs_m_chi_regions}.

These scalings can also be understood in terms of the ratio of the coupling sensitivity, $\varepsilon g'$, to the couplings required for freeze-in production, $(\varepsilon g')^{\textrm{FI}}$ (since $\bar{\sigma}_e / \bar{\sigma}_e^\text{FI} = (\varepsilon g' / (\varepsilon g')^\text{FI})^2$). For cosmologically light $m_\chi$, $(\varepsilon g')^\text{FI}$ does not depend on $m_{A'}$, while the coupling sensitivity scales as $\varepsilon g' \propto m_{A'}^0$ and $\varepsilon g' \propto m_{A'}^2$ in the light and heavy mediator limits, respectively. See App.~\ref{app:sensitivity_to_lagrangian_couplings} for further discussion of the background-free direct detection sensitivity in terms of Lagrangian couplings. 

The effect of intermediate-mass mediators is much more pronounced for a Ge target than a Si target. This can be seen in the comparison of the left and right panels in Fig.~\ref{fig:bkg_free_cs_sensitivity_mA} when $m_\chi \gtrsim 10 \, \text{MeV}$: in a Ge target, the interpolation between the light and heavy mediator limits scales approximately as $\bar{\sigma}_e/ \bar{\sigma}_e^{\textrm{FI}} \propto m_{A'}^3$, while in a Si target the scaling more rapidly asymptotes to the heavy mediator limit, $\bar{\sigma}_e/ \bar{\sigma}_e^{\textrm{FI}} \propto m_{A'}^4$. This can again be understood from the enhancement of the scattering rate in Ge due to high momentum transfers for $m_\chi \gtrsim 10$ MeV, which are less suppressed by the mediator form factor as $m_{A'}$ increases. 

So far we have restricted our discussion to a detector with an energy threshold, $\omega_{\textrm{th}}$, equal to the band gap, $E_\mathrm{g}$, of the target. Electron-based direct detection experiments typically report their energy thresholds in terms of the number of electron-hole pairs, or ionization charge, $Q$, produced in a scattering event. For an energy deposition $\omega$, this is
\begin{align}
    \label{eq:ionization}
    Q = 1 + \bigg\lfloor \frac{\omega - E_\mathrm{g}}{\epsilon} \bigg\rfloor,
\end{align}
where $\epsilon$ is the energy required for each additional electron-hole pair created. We assume $\epsilon = 3.6 \, \text{eV}\; (2.9 \, \text{eV})$ and $E_\text{g} = 1.11 \, \text{eV} \; (0.67 \, \text{eV})$ for a Si (Ge) target~\cite{10.1063/1.1656484, streetman2015solid,Essig:2015cda,Derenzo:2016fse}. An energy threshold equal to the band gap corresponds to including all electronic transitions with $Q \ge 1$; larger $\omega_\text{th}$ increases the lower bound on $Q$ according to Eq.~\eqref{eq:ionization}, and reduces the sensitivity for all DM and mediator masses. Therefore increasing $\omega_\text{th}$ will increase the ratio $\bar{\sigma}_e / \bar{\sigma}_e^{\mathrm{FI}}$ in Figs.~\ref{fig:bkg_free_cs_sensitivity_mX} and~\ref{fig:bkg_free_cs_sensitivity_mA}, thereby decreasing the experimental sensitivity.

\begin{figure*}[ht!]
    \centering
    \includegraphics[width=0.45\linewidth]{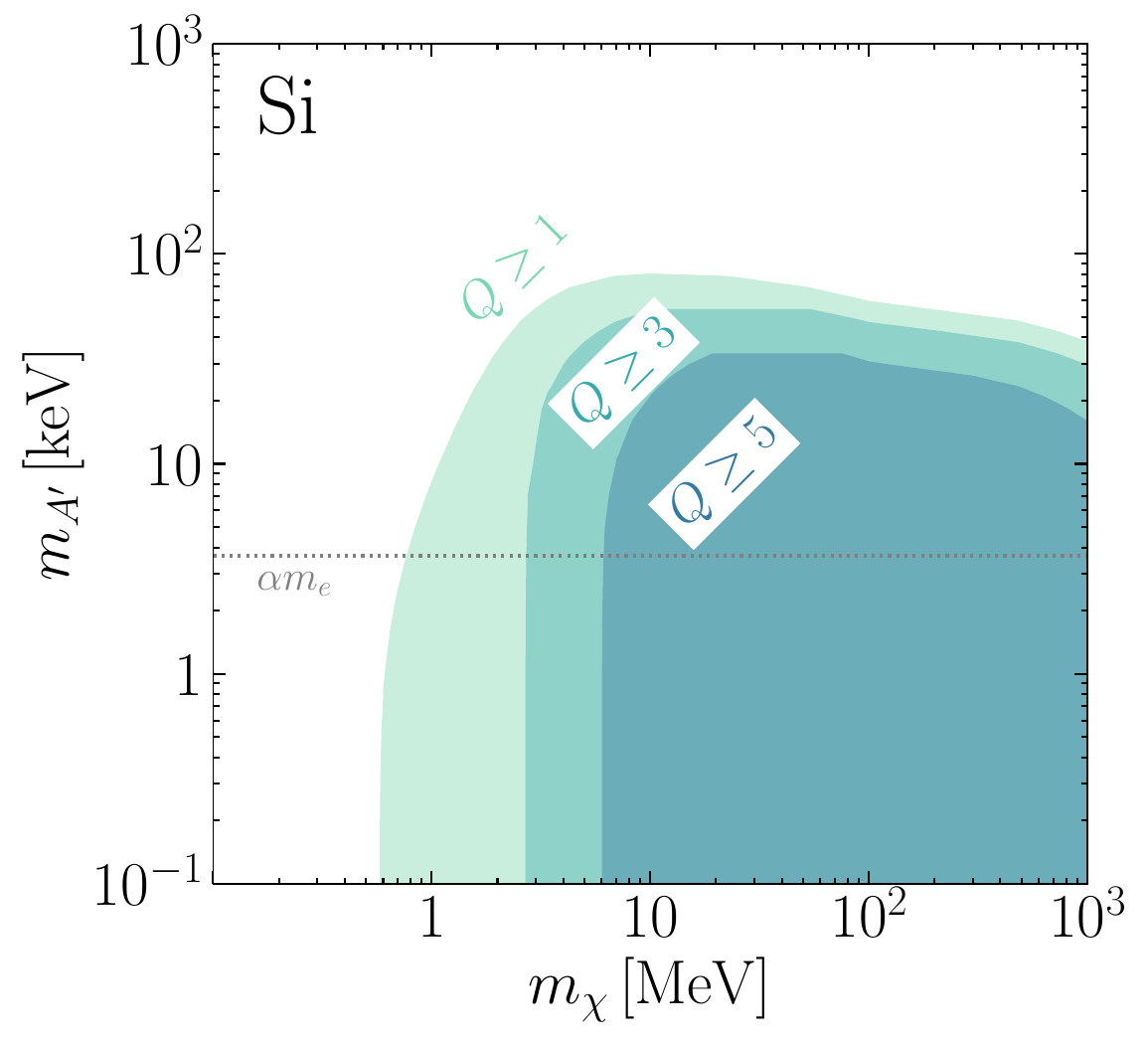}
    \includegraphics[width=0.45\linewidth]{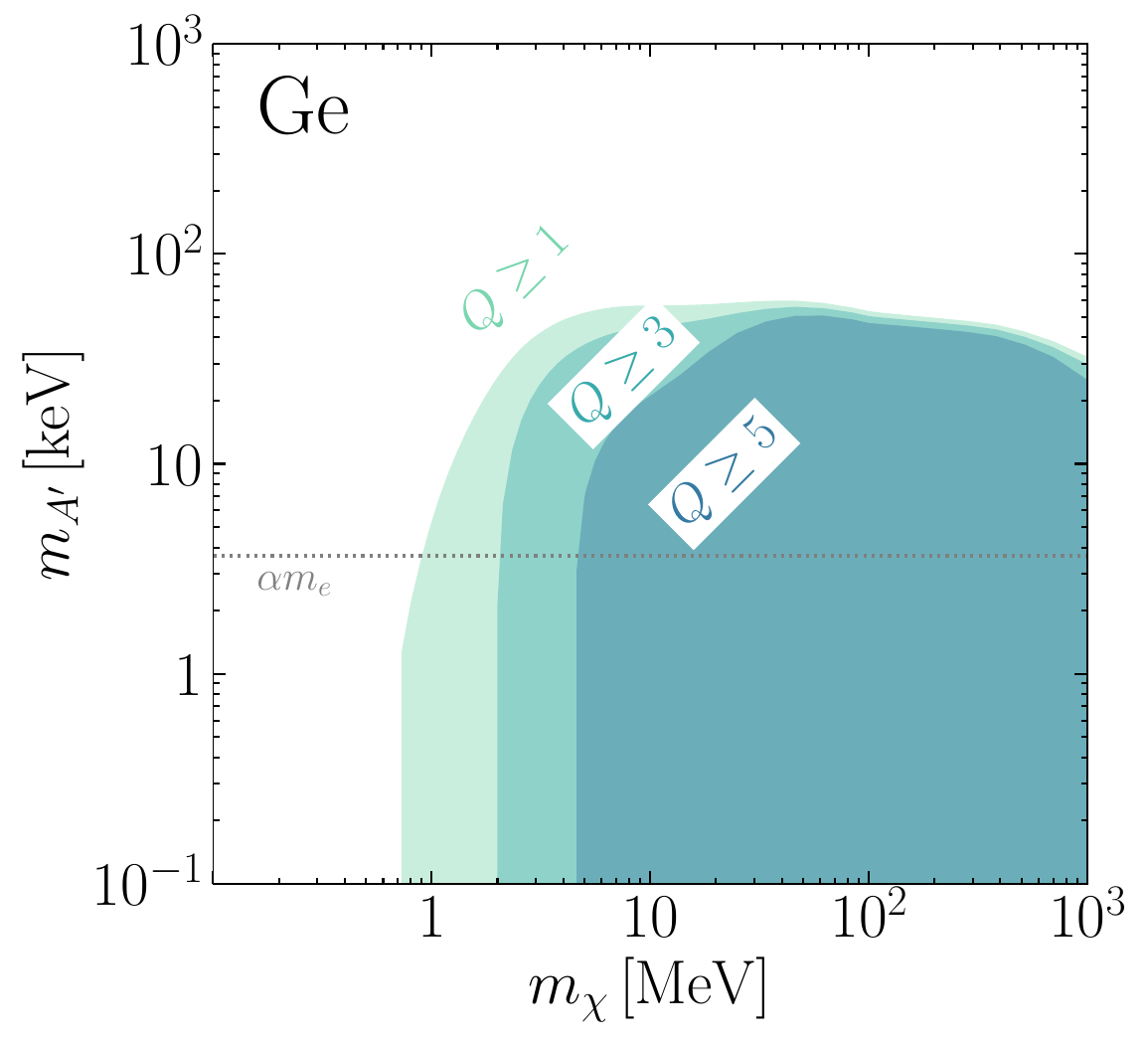}
    \caption{Projected 95\% C.L. sensitivity of background-free Si (\textbf{left}) and Ge (\textbf{right}) detectors, assuming a kg-yr exposure, to dark photon and DM masses ($m_{A'}, m_\chi$, respectively) with freeze-in sized cross sections, $\bar{\sigma}_e = \bar{\sigma}_e^\text{FI}$. Dark photon and DM masses within the shaded region are excluded for freeze-in sized cross sections ($\bar{\sigma}_e = \bar{\sigma}_e^\text{FI}$). The different colored regions correspond to different ionization charge thresholds, with higher thresholds generally resulting in reduced sensitivity to dark photon masses. For $m_\chi \gtrsim 10$ MeV in Ge, the effect of increased ionization charge threshold is minimal due the domination of the rate by large energy and momentum transfers~\cite{Griffin:2021znd,Dreyer:2023ovn,Dreyer:2026bmz}. A gray dotted line at $m_{A'}=\alpha m_e$ is shown for reference.}
    \label{fig:bkg_free_mA_vs_m_chi}
\end{figure*}

In Fig.~\ref{fig:bkg_free_mA_vs_m_chi} we present the sensitivity of a background-free experiment from a perspective complementary to Figs.~\ref{fig:bkg_free_cs_sensitivity_mX} and~\ref{fig:bkg_free_cs_sensitivity_mA}. Specifically, we show the region of $m_{A'},m_\chi$ parameter space that a background-free experiment is sensitive to with freeze-in sized cross sections, $\bar{\sigma}_e = \bar{\sigma}_e^\text{FI}$, for different ionization charge thresholds. The different colored regions correspond to ionization charge thresholds of $Q\geq 1,3,5$ (corresponding to $\omega_{\textrm{th}}=1.11, 8.31, 15.51$ eV for Si and $\omega_{\textrm{th}}=0.67, 6.47, 12.27$ eV for Ge). We expect an experiment with a given ionization charge threshold to be able to exclude freeze-in as a production mechanism for DM and dark photon masses in the shaded region. We have also shown $m_{A'} = \alpha m_e$ for reference, and to demonstrate that a detector can exclude freeze-in as a production mechanism for dark photon masses between one and two orders of magnitude larger than the conventional light mediator boundary at $m_{A'} \approx \alpha m_e$.

Increasing the threshold energy decreases the experimental sensitivity. However, for a Ge target with $m_\chi \gtrsim 10$ MeV, raising the threshold energy only marginally changes the sensitivity to $m_{A'}, m_\chi$ parameter space with freeze-in sized cross sections, $\bar{\sigma}_e = \bar{\sigma}_e^\text{FI}$, as seen in the right panel of Fig.~\ref{fig:bkg_free_mA_vs_m_chi}. Again, this is due to the dominant contributions of large $q$ to the scattering rate in Ge, and consequential peak in the DM-electron scattering rate at larger $\omega$, making the effect of the ionization charge thresholds shown negligible at these larger DM masses.

\setlength{\tabcolsep}{4.5pt}
\renewcommand{\arraystretch}{1.4}
\begin{table}[ht]
    \begin{center}
        \begin{tabular}{lcccc} \toprule
        & \multicolumn{4}{c}{\textbf{Bin}}\\
        & 2$e$ & 3$e$ & 4$e$ & 5$e$\\\cmidrule{2-5}
        Observed ($N^\text{obs}_k$) & 144 & 0 & 1 & 0 \\
        Expected Background ($N^{\textrm{b}}_k$) & 141.5 & 0.21 & 0.082 & 0.035 \\
        Collection Efficiency ($\beta_k$) &0.38 & 0.65 & 0.79 & 0.86  \\\bottomrule
        \end{tabular}
    \end{center}
    \caption{DAMIC-M~\cite{DAMIC-M:2025luv} counts data binned in number of created electron-hole pairs, $Q \,e$, where $Q$ is defined in Eq.~\eqref{eq:ionization}, and collection efficiency of the created electron hole pairs. Observed and expected background counts in each bin are summed over the different patterns. The expected background counts in each bin are assumed to be a sum of random coincidence and radioactive decays, i.e., the nuisance parameter in the background model of Ref.~\cite{DAMIC-M:2025luv} is pessimistically assumed to be one.}
    \label{tab:DAMIC_table}
\end{table}

\subsection{Projected DAMIC-M Sensitivity}
\label{sec:projected_DAMIC_sensitivity}

\begin{figure*}
    \centering
    \includegraphics[width=0.45\linewidth]{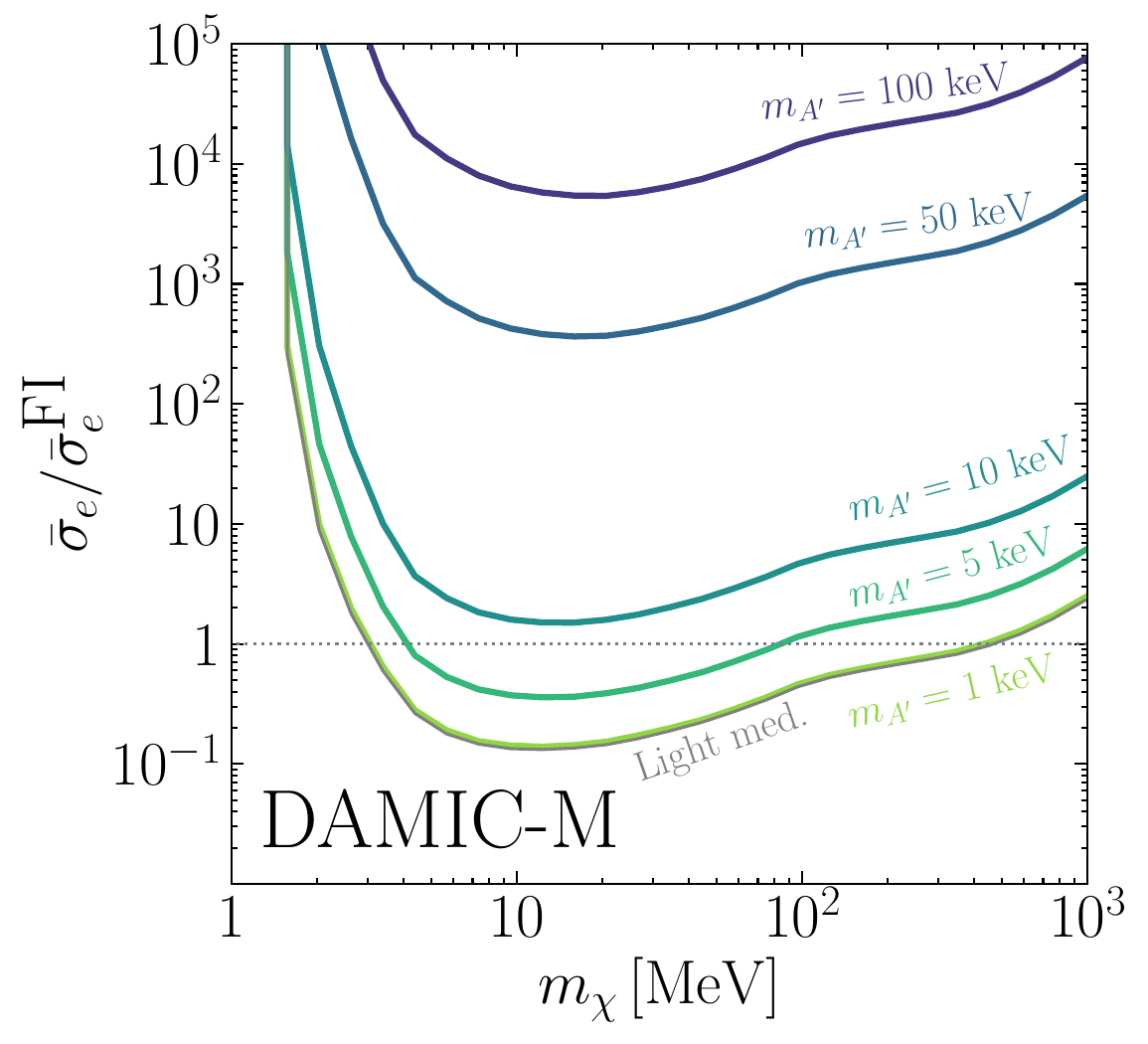}
    \includegraphics[width=0.45\linewidth]{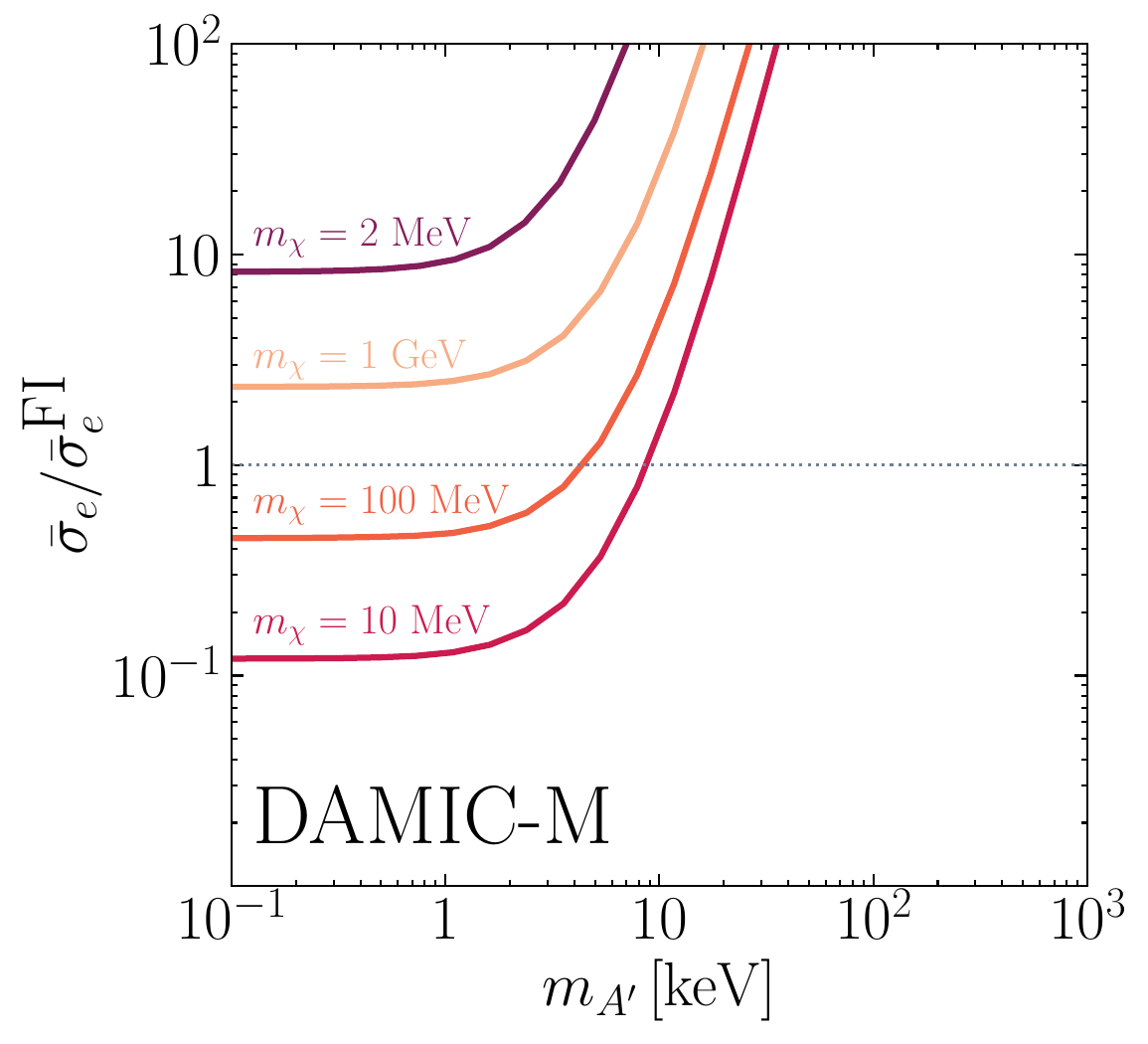}
    \caption{Ratio of projected DAMIC-M~\cite{DAMIC-M:2025luv} 90\% C.L. cross section sensitivity to the freeze-in cross section as a function of DM mass, $m_\chi$ (\textbf{left}), and dark photon mass, $m_{A'}$ (\textbf{right}), calculated using the procedure defined in Sec.~\ref{sec:projected_DAMIC_sensitivity}. DM and dark photon masses for which $\bar{\sigma}_e / \bar{\sigma}_e^\text{FI} < 1$ are expected to be ruled out by DAMIC-M, while those for which $\bar{\sigma}_e / \bar{\sigma}_e^\text{FI} > 1$ are allowed.}
    \label{fig:DAMIC_coupling_sensitivity}
\end{figure*}

We now expand our sensitivity calculation to incorporate backgrounds using the statistical treatment outlined in Refs.~\cite{Baker:1983tu,Cowan:2010js,Cirelli:2013ufw,Baxter:2021pqo,ParticleDataGroup:2024cfk}. This allows us to project the sensitivity of the DAMIC-M experiment~\cite{DAMIC-M:2025luv} using their observed and expected background counts data, which we summarize in Table~\ref{tab:DAMIC_table}. Specifically, we will determine the projected 90\% C.L. cross section sensitivity for DAMIC-M as a function of DM and dark photon mass, and compare these to the benchmark freeze-in cross section $\bar{\sigma}_e^{\textrm{FI}}$. 

Electron-based direct detection experiments typically report their results as the number of events binned in the number of electron-hole pairs created (related to energy deposited by Eq.~\eqref{eq:ionization}). For each bin, labeled by $k$, the total number of expected events is given by the sum of expected signal events, $N^{\textrm{s}}_k$, and expected background events, $N^{\textrm{b}}_k$, as
\begin{align}
    N_k = \beta_k N^{\textrm{s}}_k + N^{\textrm{b}}_k\,,
    \label{eq:expected_N_k}
\end{align}
where $\beta_k$ is a bin-dependent experimental collection efficiency. The number of expected signal events in each bin is found by calculating the predicted event rate in each bin, $R_k$ (Eq.~\eqref{eq:scattering_rate}, binned in the number of electron-hole pairs produced~\cite{Griffin:2021znd}), and multiplying by the exposure, $M \times T$, where $M$ is the target mass and $T$ is the observation time. We assume $M \times T = 1.257$ kg-day for DAMIC-M.

The likelihood of observing $\mathbf{N}^{\mathrm{obs}} = \{N_1^{\textrm{obs}},\dots,N_n^{\textrm{obs}} \}$ counts across $n$ bins given Poisson-distributed events with an expected number of events $N_k$, Eq.~\eqref{eq:expected_N_k}, is 
\begin{align}
    \label{eq:liklihood_poisson}
    L(\mathbf{N}^{\textrm{obs}}| \bar{\sigma}_e) = \prod_k \frac{N_k^{N_k^{\textrm{obs}}}}{N_k^{\textrm{obs}}!}e^{-N_k} \,.
\end{align}
The dependence on $\bar{\sigma}_e$ enters through the expected number of signal events since $N^\text{s}_k \propto \bar{\sigma}_e$. The $\bar{\sigma}_e \rightarrow 0$ limit corresponds to the no signal, or background-only, limit. 

To compare a DM model with the background-only hypothesis, we adopt the test statistic (TS),
\begin{align}
    \label{eq:TS_definition}
    \textrm{TS} \equiv -2 \ln \bigg(\frac{L(\mathbf{N}^{\textrm{obs}}| \bar{\sigma}_e)}{L(\mathbf{N}^{\textrm{obs}}| 0)}\bigg),
\end{align}
which in the limit of many events asymptotes to a $\chi^2$-distribution with a number of degrees of freedom equal to the number of free parameters in the DM model~\cite{Wilks:1935,Wilks:1938dza,Baker:1983tu}. In our case, given $m_\chi$ and $m_{A'}$, the only free parameter is $\bar{\sigma}_e$. Using the definition of the likelihood in Eq.~\eqref{eq:liklihood_poisson}, the TS can be written explicitly as
\begin{align}
    \mathrm{TS} = -2 \sum_k \bigg [N^{\mathrm{obs}}_k \ln \bigg( \frac{N_k}{N^{\mathrm{b}}_k} \bigg )
    - \beta_k N^{\textrm{s}}_k \bigg ] \, .
\end{align}
Using this form of the test statistic, a 90\% C.L. bound on the cross section is found by setting the TS equal to the appropriate $\chi^2$ value for one degree of freedom, $\text{TS} = 2.71$.

Following this statistical procedure, we compute the 90\% C.L. bounds on $\bar{\sigma}_e$, including the background counts in Table~\ref{tab:DAMIC_table}, to understand precisely which freeze-in produced DM models DAMIC-M can test. We refer to these as ``projected" limits since the statistical procedure discussed here is not exactly the same one used in Ref.~\cite{DAMIC-M:2025luv}. However, we note that this simplified statistical treatment reproduces the reported DAMIC-M 90\% C.L. cross sections to within a factor of two in the light mediator limit.

In Fig.~\ref{fig:DAMIC_coupling_sensitivity} we show the projected DAMIC-M limits relative to the freeze-in benchmark cross section, $\bar{\sigma}_e^\text{FI}$, as a function of $m_\chi$ (left) and $m_{A'}$ (right). These results are qualitatively similar to the corresponding background-free results in Fig.~\ref{fig:bkg_free_cs_sensitivity_mX} and Fig.~\ref{fig:bkg_free_cs_sensitivity_mA}, except here $\bar{\sigma}_e / \bar{\sigma}_e^\text{FI}$ is larger for all $m_\chi$ and $m_{A'}$. This reduced sensitivity is due to a lower exposure, an electron-hole creation threshold of $Q \geq2$, and the inclusion of backgrounds.

\begin{figure}[h]
    \centering
    \includegraphics[width=\linewidth]{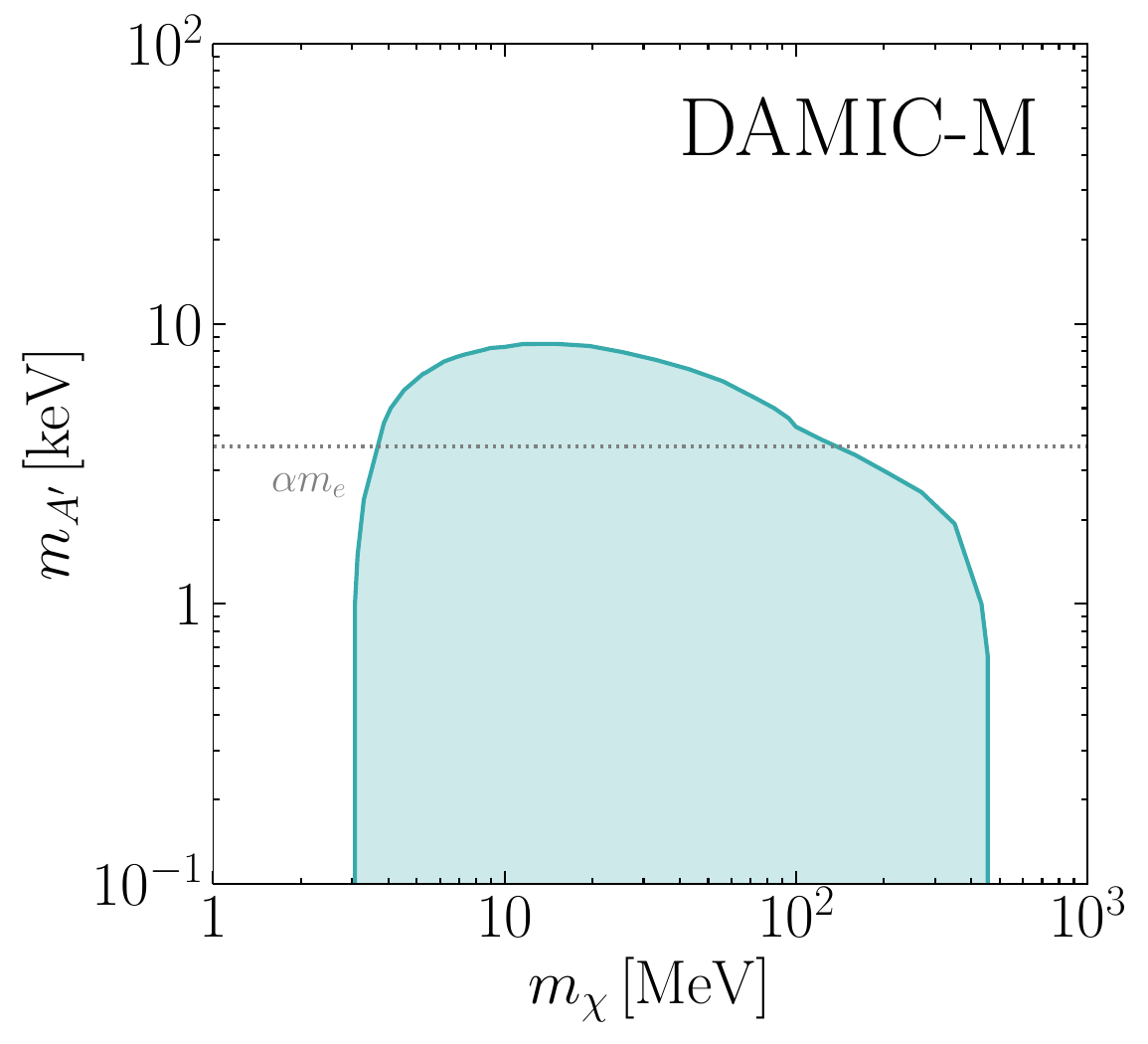}
    \caption{Projected DAMIC-M~\cite{DAMIC-M:2025luv} 90\% C.L. excluded dark photon masses, $m_{A'}$, as a function of DM mass, $m_\chi$, for cross sections that produce the DM relic abundance via freeze-in, $\bar{\sigma}_e = \bar{\sigma}_e^\text{FI}$. Models with $m_\chi, m_{A'}$ in the shaded region, and freeze-in sized cross sections ($\bar{\sigma}_e = \bar{\sigma}_e^\text{FI}$) are expected to be ruled out by DAMIC-M.}
    \label{fig:DAMIC_mA_vs_m_chi}
\end{figure}

Similar to Fig.~\ref{fig:bkg_free_mA_vs_m_chi}, in Fig.~\ref{fig:DAMIC_mA_vs_m_chi} we show the precise region of $m_\chi, m_{A'}$ parameter space that is projected to be excluded by DAMIC-M when $\bar{\sigma}_e = \bar{\sigma}_e^\text{FI}$. That is, we expect DAMIC-M can currently exclude freeze-in as a DM production mechanism for DM and dark photon masses in the shaded region. Note that the bottom left and right of the shaded region correspond to the exclusion limits for DM masses in the light mediator limit, which we find to be for $3 \, \text{MeV} \lesssim m_\chi \lesssim 460 \, \text{MeV}$. This is in close agreement with the DAMIC-M result that excludes freeze-in produced DM masses for $3.5 \, \text{MeV} \lesssim m_\chi \lesssim 490 \, \text{MeV}$ in the light mediator limit.

\section{Conclusions}
\label{sec:conclusions}

Electron-based direct detection experiments are rapidly improving their sensitivity to DM-electron scattering. Most previous calculations of DM-electron scattering rates for the Si and Ge targets used in current detectors were performed in the light or heavy mediator limits, where the mediator form factor, Eq.~\eqref{eq:exact_form_factor}, scales as $\mathcal{F} \propto 1 / q^2$ and $\mathcal{F} \propto 1$, respectively. In this paper we precisely delineate the light and heavy mediator limits, showing that they are only guaranteed to apply for a general target when $m_{A'} \ll \omega_\text{th} / v$, and $m_{A'} \gg 2\, m_\chi v$, respectively, where $\omega_\text{th}$ is the experimental threshold, $v \sim 10^{-3}$ is the DM velocity, and $m_{A'}$ is the dark photon, or mediator, mass. We define an ``intermediate-mass mediator" as a mediator whose mass is in the regime between the light and heavy limits (Eq.~\eqref{eq:regimes}, Fig.~\ref{fig:mA_vs_m_chi_regions}, Sec.~\ref{sec:intermediate_mass_mediators}). The intermediate-mass regime can be as large as three orders of magnitude for sub-GeV mass DM. We then compute the projected sensitivity of background-free electron-based direct detection experiments (Sec.~\ref{subsec:background-free_sensitivity}) and DAMIC-M~\cite{DAMIC-M:2025luv} (Sec.~\ref{sec:projected_DAMIC_sensitivity}) to light DM with intermediate-mass mediators. To compute the projected sensitivity of DAMIC-M we use a simplified statistical procedure, incorporating their expected backgrounds using Table~\ref{tab:DAMIC_table}.

Throughout we have presented the cross section sensitivity relative to the cross section needed to freeze-in the DM relic abundance via the kinetic mixing portal, $\bar{\sigma}_e^\text{FI}$. This allows for a straightforward determination of whether a point in $m_\chi, m_{A'}$ parameter space is detectable, or excluded, with freeze-in sized couplings ($\bar{\sigma}_e = \bar{\sigma}_e^\text{FI}$), which we illustrate in Figs.~\ref{fig:bkg_free_mA_vs_m_chi} and~\ref{fig:DAMIC_mA_vs_m_chi}. We find that for MeV-GeV mass DM, a background-free, kg-yr exposure, experiment would be sensitive to $m_{A'} \lesssim 100 \, \text{keV}$ and we project DAMIC-M is sensitive to $m_{A'} \lesssim 10 \, \text{keV}$, for freeze-in sized couplings. This precisely demarcates the sensitivity of direct detection experiments to freeze-in produced DM interacting with the Standard Model through a kinetically-mixed dark photon (Sec.~\ref{subsec:interaction_lagrangian}) as a function of DM and dark photon mass. We note that some of the parameter space with intermediate-mass dark photons is also constrained by both stellar energy loss via dark photons directly~\cite{Fabbrichesi:2020wbt,Caputo:2025avc,Caputo:2026pdw} and self-interacting DM (SIDM) limits on the dark gauge coupling~\cite{Knapen:2017xzo}. However, these bounds are not absolute: density-dependent mechanisms can suppress stellar energy loss~\cite{DeRocco:2020xdt,Chakraborty:2020vec}, and the SIDM constraints are relaxed for DM sub-components~\cite{Cyr-Racine:2013fsa,Knapen:2017xzo}. The direct detection limits derived here are therefore complementary to these other astrophysical probes.

An important next step will be to incorporate the \textit{exact} analysis chain of DAMIC-M, and other ongoing direct detection experiments when results are ready, to understand the exact experimental sensitivity to intermediate-mass mediators. To aid in this effort we make the results of the DM-electron scattering rate calculations in the intermediate-mass mediator regime publicly available at Ref.~\cite{stratman_2026_19805613}. Additionally, it would be interesting to compare the results derived here with \EXCEEDDM~to the recently released QCDark2~\cite{Dreyer:2026bmz} which includes local field effects. While the cross section sensitivity in light and heavy mediator limits is qualitatively similar, local field effects incur non-trivial momentum dependence in the target response which may be important when considering intermediate-mass mediators.

Lastly, most of our discussion surrounding the intermediate-mass mediator regime can be straightforwardly extended to the next generation of direct detection experiments with $\mathcal{O}(\text{meV})$ thresholds~\cite{Lin:2019uvt,Kahn:2021ttr,Zurek:2024qfm}. In fact, for these experiments the intermediate-mass regime is even larger, extending from $\text{eV} \lesssim m_{A'} \lesssim \text{MeV}$ for keV-GeV mass DM, and it will be important to understand what part of this parameter space future single-phonon~\cite{Schutz:2016tid,Knapen:2017ekk,Griffin:2019mvc,TESSERACT:2025tfw}, single-magnon~\cite{Trickle:2019ovy}, and small-gap semiconductor~\cite{Abbamonte:2025guf} detectors can probe.

\acknowledgments
We would like to thank Rouven Essig, Megan Hott, Gordan Krnjaic, Robert McGehee, and Jessie Shelton for useful discussion. This work made use of the Illinois Campus Cluster, a computing resource that is operated by the Illinois Campus Cluster Program (ICCP) in conjunction with the National Center for Supercomputing Applications (NCSA) and which is supported by funds from the University of Illinois Urbana-Champaign.


\appendix

\begin{figure*}[ht!]
    \centering
    \includegraphics[width=0.45\linewidth]{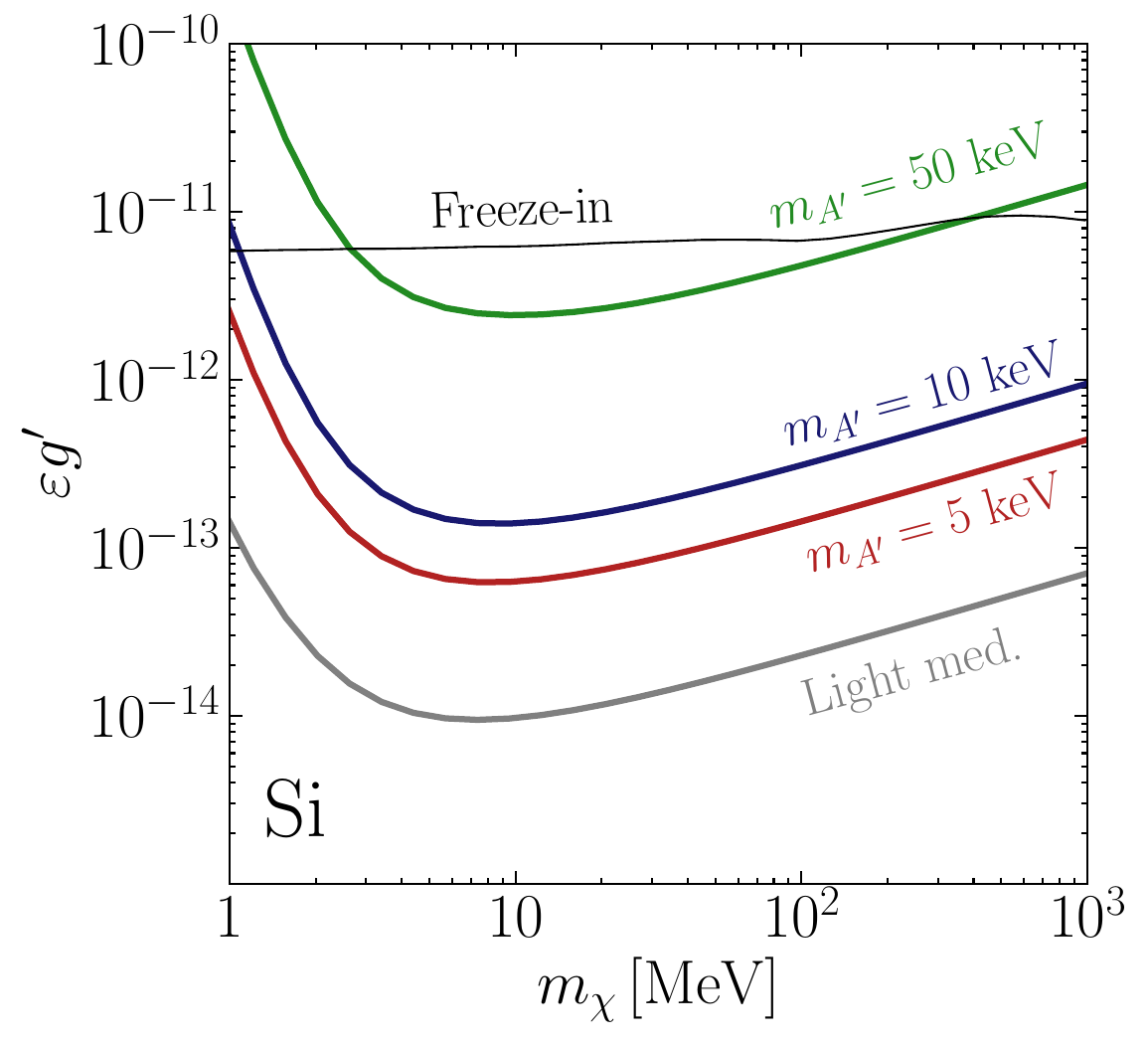}
    \includegraphics[width=0.45\linewidth]{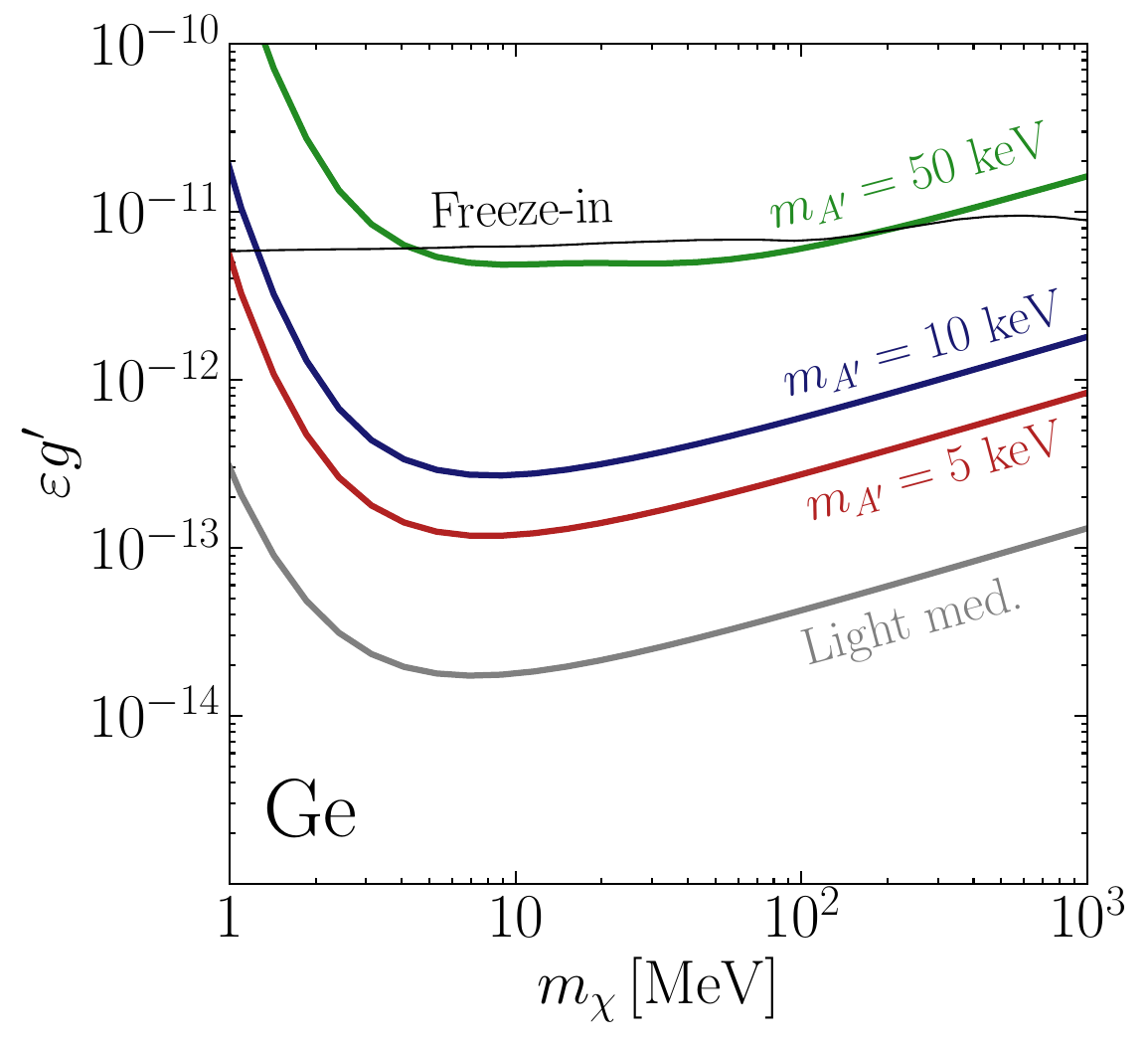}
    \caption{Projected 95\% C.L. sensitivity to the Lagrangian couplings, $\varepsilon g'$ (Eq.~\eqref{eq:L_int}), of background-free Si (\textbf{left}) and Ge (\textbf{right}) detectors, assuming an ionization threshold $Q \geq 1$ and kg-yr exposure, for a variety of dark photon masses, $m_{A'}$. This figure is equivalent to the left panels of Fig.~\ref{fig:bkg_free_cs_sensitivity_mX}, upon converting the cross section sensitivity to the corresponding product of the kinetic mixing parameter, $\varepsilon$, and $U(1)'$ gauge coupling, $g'$, using Eq.~\eqref{eq:cross_section_definition}. Here, the freeze-in couplings (thin black line) are independent of the dark photon mass.}
    \label{fig:couplings_mX}
\end{figure*}

\section{Reference Cross Section Definitions}
\label{app:reference_cross_section_definitions}

Previous works have used a variety of definitions for the reference DM-electron cross section, $\bar{\sigma}_e$. In this appendix we provide equations to convert between the conventions. One common convention is to use the Standard Model fine structure constant, or the dark sector fine structure constant, $\alpha' = g'^2 / 4 \pi$, instead of the $U(1)$ or $U(1)'$ gauge couplings,
\begin{align}
    \bar{\sigma}_e = \mu_{\chi e}^2 \frac{4 \, \alpha \, g'^2 \, \varepsilon^2 }{\left( \alpha^2 m_e^2 + m_{A'}^2 \right)^2} = \mu_{\chi e}^2 \frac{16 \pi \, \alpha \, \alpha' \, \varepsilon^2 }{\left( \alpha^2 m_e^2 + m_{A'}^2 \right)^2} \, .
\end{align}
The first convention is followed in Ref.~\cite{Essig:2015cda}, and the second in Refs.~\cite{Essig:2011nj,Knapen:2017xzo,Lin:2019uvt,Zurek:2024qfm,Krnjaic:2025noj}.

Another common convention is to combine the $U(1)'$ gauge coupling and kinetic mixing parameter in to a single coupling, $\kappa$, 
\begin{align}
    \kappa \equiv \frac{\varepsilon g'}{e} = \varepsilon \sqrt{ \frac{\alpha'}{\alpha} } \, ,
\end{align}
which alone determines freeze-in production and direct detection detectability. Additionally, note that $\kappa$ is also equal to the millicharge parameter, $Q$, which is sometimes used in the limit of a light dark photon~\cite{Dvorkin:2019zdi,Lin:2019uvt}, as in this limit one can equivalently redefine $A'$ to remove the kinetic mixing term in Eq.~\eqref{eq:L}. In terms of $\kappa$ the reference DM-electron cross section is,
\begin{align}
    \bar{\sigma}_e = \frac{\mu_{\chi e}^2}{\pi} \frac{e^4 \kappa^2}{\left( \alpha^2 m_e^2 + m_{A'}^2 \right)^2} = \mu_{\chi e}^2 \frac{16 \pi \, \alpha^2 \kappa^2}{\left( \alpha^2 m_e^2 + m_{A'}^2 \right)^2}\, ,
\end{align}
which is used in Refs.~\cite{Bhattiprolu:2023akk,Bhattiprolu:2024dmh}. Additionally, note that $\varepsilon$ here (Eq.~\eqref{eq:L_int}) is referred to as $\hat{\epsilon}$ in Ref.~\cite{Chu:2011be}, since $\epsilon$ in Ref.~\cite{Chu:2011be} (or $\epsilon_Y$ in Ref.~\cite{Bhattiprolu:2024dmh})) is reserved for the parameter determining the kinetic hypercharge mixing, relevant at energies above the electroweak scale. The relationship between these epsilons relative to Ref.~\cite{Chu:2011be} is $\varepsilon = \hat{\epsilon} = \epsilon \cos{\theta_W}$, where $\theta_W$ is the Weinberg angle.

Lastly, the simplest convention, useful when thinking in the context of simplified DM models, is to start from the interaction Lagrangian,
\begin{align}
    \mathcal{L} \supset -A'_\mu \left( g_\chi \, \bar{\chi} \gamma^\mu \chi + g_e \, \bar{e} \gamma^\mu e  \right) 
    \label{eq:L_int_app}
\end{align}
which just has two couplings, $g_\chi$, $g_e$, which parameterize the DM-dark photon and electron-dark photon interaction strength, respectively. Comparing Eq.~\eqref{eq:L_int_app} to Eq.~\eqref{eq:L_int} we see that $g_\chi, g_e$ are related to the parameters used here as,
\begin{align}
    g_\chi = g'~~~,~~~g_e = -\varepsilon e \, .
\end{align}
The reference DM-electron cross section in this convention is then
\begin{align}
   \bar{\sigma}_e = \frac{\mu_{\chi e}^2}{\pi} \frac{g_\chi^2 g_e^2}{\left( \alpha^2 m_e^2 + m_{A'}^2 \right)^2} \, ,
\end{align}
which is used in Refs.~\cite{Knapen:2021run,Hochberg:2021pkt,Dreyer:2026bmz}.

To summarize, the comparison of the reference DM-electron cross sections used throughout the literature is,
\begin{align}
    \bar{\sigma}_e & \equiv \frac{\mu_{\chi e}^2}{\pi} \frac{g'^2 \, (\varepsilon e)^2}{\left( \alpha^2 m_e^2 + m_{A'}^2 \right)^2} = \mu_{\chi e}^2 \frac{4 \, \alpha \, g'^2 \, \varepsilon^2 }{\left( \alpha^2 m_e^2 + m_{A'}^2 \right)^2} \nonumber\\
    & = \mu_{\chi e}^2 \frac{16 \pi \, \alpha \, \alpha' \, \varepsilon^2 }{\left( \alpha^2 m_e^2 + m_{A'}^2 \right)^2} = \frac{\mu_{\chi e}^2}{\pi} \frac{e^4 \kappa^2}{\left( \alpha^2 m_e^2 + m_{A'}^2 \right)^2} \nonumber\\
    & = \mu_{\chi e}^2 \frac{16 \pi \, \alpha^2 \kappa^2}{\left( \alpha^2 m_e^2 + m_{A'}^2 \right)^2} = \frac{\mu_{\chi e}^2}{\pi} \frac{g_\chi^2 g_e^2}{\left( \alpha^2 m_e^2 + m_{A'}^2 \right)^2} \, ,
\end{align}

\begin{align}
    \kappa \equiv \frac{\varepsilon g'}{e} &= \varepsilon \sqrt{ \alpha' / \alpha }~~~,~~~g_\chi = g'~~~,~~~g_e = -\varepsilon e \, .
    \label{eq:big_comparison}
\end{align}

\section{Sensitivity to Lagrangian Couplings}
\label{app:sensitivity_to_lagrangian_couplings}

Fig.~\ref{fig:couplings_mX} compares the background-free sensitivity of Si and Ge targets against the benchmark freeze-in target in terms of the Lagrangian couplings, Eq.~\eqref{eq:L}; specifically the kinetic mixing parameter, $\varepsilon$, and $U(1)'$ gauge coupling, $g'$. In this parameterization, for each $m_\chi$, the freeze-in couplings for cosmologically light dark photons have one specific value, but the sensitivity of electron-based direct detection experiments changes dramatically as $m_{A'}$ changes. We illustrate this behavior in Fig.~\ref{fig:couplings_mX}, which contains the same information as the left panels of Fig.~\ref{fig:bkg_free_cs_sensitivity_mX}, but is converted to $\varepsilon g'$ from $\bar{\sigma}_e$ using Eq.~\eqref{eq:cross_section_definition}.

\bibliographystyle{utphys3}
\bibliography{references}

\end{document}